\documentclass[onecollarge]{svjour2}       
\smartqed  

\usepackage[dvipdfmx]{graphicx}

\usepackage{amsmath, amssymb}
\usepackage{amsfonts}
\usepackage{multicol}
\usepackage{braket}

\usepackage{ascmac}
\usepackage{fancybox}
\usepackage{comment}

\usepackage{cite}

\newcommand{\eq}[1]{\begin{equation} #1 \end{equation}}
\newcommand{\eqa}[2]{\begin{equation} #1 \label{#2} \end{equation}}
\newcommand{\balign}[1]{\begin{align} #1 \end{align}}

\newcommand{\fn}{\footnote}

\newcommand{\todayd}{\the\year/\the\month/\the\day}

\newcommand{\bib}{\bibitem}

\newcommand{\up}{\uparrow}

\newcommand{\lr}{\leftrightarrow}

\newcommand{\lb}{\label}
\newcommand{\nt}{\notag}

\newcommand{\eref}[1]{Eq.~\eqref{#1}}

\newcommand{\sref}[1]{Sec.~\ref{s:#1}}
\newcommand{\cref}[1]{Chap.~\ref{c:#1}}

\newcommand{\bel}{\begin{easylist}}
\newcommand{\eel}{\end{easylist}}

\newcommand{\be}[1]{\begin{enumerate} #1 \end{enumerate}}

\def \({\left(}
\def \){\right)}

\def \[{\left[}
\def \]{\right]}
\newcommand{\abs}[1]{\left|#1\right|}

\newcommand{\sumtwo}[2]%
{\mathop{\sum_{#1}}_{#2}}
\newcommand{\sumthree}[3]%
{\mathop{\mathop{\sum_{#1}}_{#2}}_{#3}}
\newcommand{\sumfour}[4]%
{\mathop{\mathop{\mathop{\sum_{#1}}_{#2}}_{#3}}_{#4}} 
\newcommand{\prodtwo}[2]%
{\mathop{\prod_{#1}}_{#2}}
\newcommand{\mintwo}[2]%
{\mathop{\min_{#1}}_{#2}}
\newcommand{\maxtwo}[2]%
{\mathop{\max_{#1}}_{#2}}
\newcommand{\maxthree}[3]%
{\mathop{\mathop{\max_{#1}}_{#2}}_{#3}}
\newcommand{\limtwo}[2]%
{\mathop{\lim_{#1}}_{#2}}
\newcommand{\suptwo}[2]%
{\mathop{\sup_{#1}}_{#2}}
\newcommand{\supthree}[3]%
{\mathop{\mathop{\sup_{#1}}_{#2}}_{#3}}
\newcommand{\supfour}[4]%
{\mathop{\mathop{\mathop{\sup_{#1}}_{#2}}_{#3}}_{#4}} 
\newcommand{\inftwo}[2]%
{\mathop{\inf_{#1}}_{#2}}
\newcommand{\infthree}[3]%
{\mathop{\mathop{\inf_{#1}}_{#2}}_{#3}}
\newcommand{\inffour}[4]%
{\mathop{\mathop{\mathop{\inf_{#1}}_{#2}}_{#3}}_{#4}} 

\newcommand\calP{{\cal P}}

\newcommand{\bsA}{\boldsymbol{A}}
\newcommand{\bsB}{\boldsymbol{B}}
\newcommand{\bsC}{\boldsymbol{C}}
\newcommand{\bsD}{\boldsymbol{D}}

\newcommand{\bbC}{\mathbb{C}}

\newcommand{\bbR}{\mathbb{R}}

\newcommand{\ba}[2]{
\begin{array}{#1}
#2
\end{array}
}

\def\x{\overline{X}}
\def\y{\overline{Y}}
\def\z{\overline{Z}}
\def\a{\overline{A}}
\def\b{\overline{B}}

\def\d{\overline{D}}

\def\tx{\widetilde{X}}
\def\ty{\widetilde{Y}}
\def\tz{\widetilde{Z}}
\def\brz{\overset{\ zz}{|}}
\def\blz{\overset{zz \ }{|}}
\def\hrz{\underset{\ zz}{\up}}
\def\hlz{\underset{zz \ }{\up}}

\def\tb{\widetilde{B}}
\def\tL{\widetilde{L}}
\def\tR{\widetilde{R}}

\def\bz{\overset{Z}{|}}
\def\hz{\underset{Z}{\up}}

\newcommand{\qh}{$[Q, H]$ }
\def\i{{\rm i}}

\newcommand{\lplus}{\overset{\to}{+}}
\newcommand{\rplus}{\overset{\leftarrow}{+}}
\newcommand{\threeskip}{\ \ }

\newcommand{\cX}{c_{\rm XYZh}}
\newcommand{\cN}{c_{\rm NNN}}

\def\rnum#1{\resizebox{0.5em}{\height}{\expandafter{\romannumeral #1}}}
\def\Rnum#1{\resizebox{0.5em}{\height}{\uppercase\expandafter{\romannumeral #1}}}

\newcommand{\bthm}[1]{\begin{theorem} #1 \end{theorem}}
\newcommand{\blm}[1]{\begin{lemma} #1 \end{lemma}}
\newcommand{\bdf}[1]{\begin{definition} #1 \end{definition}}

\begin{document}

\title{Absence of local conserved quantity in the Heisenberg model with next-nearest-neighbor interaction}


\author{Naoto Shiraishi }


\institute{Naoto Shiraishi  \at
              Faculty of arts and sciences, University of Tokyo, Komaba 3-8-1, Meguro-ku, Tokyo, Japan \\
             \email{shiraishi@phys.c.u-tokyo.ac.jp}           
          }

\date{Received: date / Accepted: date}

\maketitle

\begin{abstract}
We rigorously prove that the $S=1/2$ anisotropic Heisenberg chain (XYZ chain) with next-nearest-neighbor interaction, which is anticipated to be non-integrable, is indeed non-integrable in the sense that this system has no nontrivial local conserved quantity.
Our result covers some important models including the Majumdar-Ghosh model, the Shastry-Sutherland model, and many other zigzag spin chains as special cases.
These models are shown to be non-integrable while they have some solvable energy eigenstates.
In addition to this result, we provide a pedagogical review of the proof of non-integrability of the $S=1/2$ XYZ chain with Z magnetic field, whose proof technique is employed in our result.

\end{abstract}

\setcounter{tocdepth}{2}
\tableofcontents

\section{Introduction}\label{intro}
Integrable systems, or exactly solvable systems, are special quantum many-body systems whose energy eigenstates can be computed exactly~\cite{JM, Fad, Bax, Tak}.
The structure behind solvability is the existence of (infinitely many) local conserved quantities, which decomposes the Hilbert space and helps to construct energy eigenstates~\cite{DG, FF, GM94, GM95, GM95-2, KBI}.
An established method to obtain local conserved quantities in integrable systems is the quantum inverse scattering method~\cite{Fad}, which algebraically reproduces the solutions obtained by the Bethe ansatz.
The quantum inverse scattering method has revealed the integrability of the Heisenberg model, the XXZ model, and various more complex systems~\cite{ABF, AKW, OY}.

In contrast to the aforementioned deep understanding of integrable systems, very few studies have tackled non-integrable systems.
Here, we identify non-integrability to the absence of nontrivial local conserved quantity.
Non-integrability is a necessary condition for the application of the Kubo formula in the linear response theory~\cite{Kub, SF, SK, Suz} and for the existence of normal transport~\cite{Zot} and thermalization phenomena~\cite{Caz, RDYO, Lan, EF}, which suggests the importance of clarifying non-integrability of quantum many-body systems.
In spite of its importance and expected ubiquitousness of non-integrable systems, quantum non-integrability has not been studied from an analytical viewpoint for a long time (A notable exception is \cite{GM95-2}, while its attempt is heuristic and relies on some plausible assumptions).

Recently, the author invented a method to prove the non-integrability of quantum many-body systems, with which the XYZ model with Z magnetic field is proven to be non-integrable~\cite{Shi19}.
By applying this method, the mixed-field Ising chain~\cite{Chi24} and the PXP model~\cite{PL24} were also shown to be non-integrable.
However, these three applications are presented as craftsmanship, and the potential power and the general structure of this method have not yet been clarified.

In this paper, we prove the non-integrability of the Heisenberg model and the XYZ model with next-nearest-neighbor interactions.
This model is also called as {\it zigzag spin ladders}, where we regard the next-nearest-neighbor interaction on odd sites (1,3,5,$\ldots$) and even sites (2,4,6,$\ldots$) as two parallel lines and the nearest-neighbor interactions as ladders.
We show that if all of the next-nearest-neighbor interactions (X, Y, and Z) are nonzero and one of the nearest-neighbor interactions with X, Y, and Z is nonzero, then this system has no nontrivial local conserved quantity.
Our setup includes an important model, the Majumdar-Ghosh model~\cite{MG}, as its special case.
The Majumdar-Ghosh model is a famous frustration-free system, whose ground state and several excited energy eigenstates can be solved exactly~\cite{CM82, CEM, Ber18}.
In addition to this model, various frustration-free zigzag spin chains~\cite{Ham88, Bat09, BS12, SH24} are also special cases of our setup.
By combining our result and the frustration-free property of these models, the Majumdar-Ghosh model and other zigzag spin chain models mentioned above are rigorously proven to be interesting models where several energy eigenstates are solvable but most of energy eigenstates are unsolvable, at least by the quantum inverse scattering method.
With a slight extension, a quantum chain model proposed by Shastry and Sutherland~\cite{SS81} is also shown to have no local conserved quantity.


This paper is organized as follows.
In \sref{claim}, we rigorously define the local conserved quantity and its absence and present our main theorems.
We claim that the XYZ chain with Z magnetic field and the Heisenberg chain with next-nearest-neighbor interaction have no local conserved quantity.
Although the non-integrability of the former has already been shown in \cite{Shi19}, in this paper, we provide a pedagogical explanation of this proof since this proof serves as a prototype of other applications.
In \sref{proof-idea}, we explain the proof method for the absence of local conserved quantity:
Expanding a candidate of a local conserved quantity by the basis with the Pauli matrices and the identity operator, we demonstrate that all of the coefficients of these operators are zero by using the fact that it commutes with the Hamiltonian.
In \sref{symbol1}, we introduce several useful symbols employed in our proofs.

In \sref{XYZh}, we review the proof of the absence of local conserved quantity in the XYZ chain with Z magnetic field.
Since some of its proof idea is directly extended to our main result in \sref{NNN}, we explain this proof in a pedagogical and systematic way.
In \sref{k=3demo}, we treat the simplest case as a simple demonstration and show that all the remaining coefficients are zero.
In \sref{XYZh-step1}, we restrict a possible form of nonzero coefficients.
In \sref{symbol2} we introduce several symbols, and using them, in \sref{XYZh-step2} we demonstrate that all the remaining coefficients are zero, which completes our proof.

\sref{NNN} is our main part, where we prove the absence of local conserved quantity in the Heisenberg chain with next-nearest-neighbor interaction.
In \sref{NNN-step1} and \sref{NNN-step2}, we restrict a possible form of nonzero coefficients, along with a similar line to \sref{XYZh-step1}.
In \sref{NNN-step3}, with using symbols introduced in \sref{symbol2}, we demonstrate that all the remaining coefficients have zero coefficients, which completes our proof.
We slightly extend our result in \sref{extend-NNN}, with which the Shastry-Sutherland model (a Heisenberg-type quantum chain model, not the famous two-dimensional model)  is covered.

\section{Problem and general strategy}\lb{s:problem}

\subsection{Claim}\lb{s:claim}

This paper aims to prove the absence of local conserved quantity, which we employ as the definition of the quantum non-integrability in this paper, in two spin models which are considered to be non-integrable.
One is the standard $S=1/2$ XYZ spin chain on $L$ sites with a magnetic field in $z$-direction with the periodic boundary condition.
By denoting by $X$, $Y$, $Z$ the Pauli matrices $\sigma^x$, $\sigma^y$, $\sigma^z$, the Hamiltonian is expressed as
\eqa{
H=-\sum_{i=1}^L [ J^X X_i X_{i+1}+J^Y Y_iY_{i+1}+J^ZZ_iZ_{i+1}] -\sum_{i=1}^L h Z_i
}{XYZ+h}
with setting all the coupling constants $J^X$, $J^Y$, $J^Z$ nonzero.
Here, we identify site $L+1$ to site 1, meaning the periodic boundary condition.
We call this model as {\it XYZ+h model} in short.
The non-integrability of XYZ+h model with $J_X\neq J_Y$ and $h\neq 0$ is shown in \cite{Shi19}.
We review this result in \sref{XYZh}.

The other model we treat is the $S=1/2$ XYZ model with next-nearest-neighbor interaction, which is the main subject of this paper.
The Hamiltonian of this model is expanded as
\eq{
H=\sum_{i=1}^L  [ J_1^X X_i X_{i+1}+J_1^Y Y_iY_{i+1}+J_1^ZZ_iZ_{i+1}] + \sum_{i=1}^L[ J_2^X X_i X_{i+2}+J_2^Y Y_iY_{i+2}+J_2^ZZ_iZ_{i+2}] ,
}
where, we identify sites $L+1$ and $L+2$ to sites 1 and 2, implying the periodic boundary condition.
We call this model as {\it NNN-XYZ model} in short.
This model is also called a zigzag spin chain, which has been investigated in the context of frustration-free systems~\cite{Ham88, Bat09, BS12, SH24}.
If $J_1^a$ and $J_2^a$ ($a\in \{X, Y, Z\}$) do not depend on the direction $a$, this Hamiltonian is isotropic and reduces to the Heisenberg-type interaction.
In particular, the Majumdar-Ghosh model~\cite{MG} is included as a special case.
We suppose that all of $J_2^X$, $J_2^Y$, $J_2^Z$ and one of  $J_1^X$, $J_1^Y$, $J_1^Z$ are nonzero.
In the following, we set $J_1^Z$ nonzero.
We treat NNN-XYZ model in \sref{NNN}, where the proof technique presented in \sref{XYZh} is essential.

\bigskip

In order to state our claim rigorously, we first clarify the notion of the locality of operators.

\bdf{
An operator $C$ is a {\it $k$-support operator} if its contiguous support is among $k$ sites.
}

Let us see several examples.
With promising that the subscript of an operator represents the site it acts, an operator $X_4Y_5Z_6$ is a 3-support operator and $X_2X_5$ is a 4-support operator, since the contiguous support of the latter is $\{2,3,4,5\}$.
In case without confusion, we call the shift-sum of $k$-support operators, e.g., $\sum_i X_iY_{i+1}X_{i+2}$, also simply as a {\it $k$-support operator}.

We express a sequence of $l$ operators with $A\in \{ X,Y,Z,I\}$ starting from site $i$ to site $i+l-1$ by a shorthand symbol $\bsA^l_i:=A^1_iA^2_{i+1}\cdots A^l_{i+l-1}$.
We promise that the first and the last operators $A^1$ and $A^l$ take one of the Pauli operators ($X$, $Y$, or $Z$), not an identity operator $I$, while other operators $A^2,\cdots ,A^{l-1}$ are one of $\{ X,Y,Z,I\}$.
We denote a set of such operator sequences $A^1A^2\cdots A^l$ by $\calP^l$.
Using these symbols, a candidate of a shift-invariant local conserved quantity can be expressed as
\eqa{
Q=\sum_{l=1}^{k} \sum_{\bsA^l\in \calP^l} \sum_{i=1}^L q_{\bsA^l}\bsA^l_i
}{Qform}
with coefficients $q_{\bsA^l}\in \bbR$.
The sum of $\bsA^l$ runs over all possible $9\times 4^{l-2}$ sequences of operators from $XX\cdots XX$ to $ZI\cdots IZ$.
The Pauli matrices and the identity span the space of $2\times2$ Hermitian matrices, which confirms that the above form covers all possible shift invariant quantities whose contiguous support of summand is less than or equal to $k$.

\bdf{
An operator $Q$ in the form of \eref{Qform} is a {\it $k$-support conserved quantity} if (i) $Q$ is conserved in the sense that $[Q,H]=0$, and (ii) one of $q_{\bsA ^k}$ is nonzero.
}

Conventionally, a {\it local} conserved quantity refers to a $k$-support conserved quantity with $k=O(1)$ with respect to the system size $L$.
Our main results exclude a much larger class of conserved quantities, including some $k=O(L)$ cases.

\bthm{
The XYZ+h model with $J^X\neq J^Y$ and $h\neq 0$ has no $k$-support conserved quantity with  $3\leq k\leq L/2$.
}

\bthm{
The NNN-XYZ model with $J_2^X, J_2^Y, J_2^Z\neq 0$ and $J_1^Z\neq 0$ has no $k$-support conserved quantity with $4\leq k\leq L/2-1$.
}

These results demonstrate that all nontrivial conserved quantities in these models are highly nonlocal, implying that the quantum inverse scattering method never solves these models.
We note that the upper bound of $k$ is almost tight because the square of the Hamiltonian, $H^2$, is a $L/2+2$-support conserved quantity in the XYZ+h model, and is a $L/2+3$-support conserved quantity in the NNN-XYZ model.
We also note that the system has some trivial conserved quantities including the Hamiltonian itself and, in the case of symmetry, a magnetic field.
They are $k$-support conserved quantities with $k$ less than the conditions in the above theorems.

We emphasize that although we restrict possible conserved quantities in the shift-invariant form, this restriction does not decrease the generality of our result.
We explain its intuitive reason below and rigorously justify it in the Appendix.
Suppose that $Q$ is not shift invariant.
Then, by defining $T$ as the shift operator by one site, $\sum_{j=1}^L T^{j} QT^{-j}$ is also a conserved quantity and is now shift-invariant.
Hence, it suffices to treat shift-invariant conserved quantities.

\subsection{Proof idea}\lb{s:proof-idea}

Let $\mu$ be the size of the maximum contiguous support of $H$, which is two in the XYZ+h model and three in the NNN-XYZ model.
We first notice that the commutator of a $k$-support operator $Q$ and $H$ is an at most $k+\mu-1$-support operator, which guarantees the expansion as
\eqa{
[Q,H]=\sum_{l=1}^{k+\mu-1}\sum_{\bsB^l\in \calP^l} \sum_{i=1}^L r_{\bsB^l} \bsB^l_i.
}{QH}
The conservation of $Q$ implies $r_{\bsB^l}=0$ for any $\bsB^l$, which leads to many constraints (linear relations) on $q_{\bsA}$ in \eref{Qform} by comparing both sides of \eref{QH}.
Our goal is to show that these linear relations do not have nontrivial solutions except for $q_{\bsA ^k}=0$ for all $\bsA^k$, which means that $Q$ cannot be a $k$-support conserved quantity.

Our proof consists of two steps.
\begin{enumerate}
\renewcommand{\labelenumi}{\alph{enumi}.}
\item We examine the condition $r_{\bsB^{l}}=0$ for all $\bsB^{l}$ from $l=K+\mu-1$ to $l=K+1$, and show that the coefficients of $\bsA^k$ except those in a specific form (i.e., doubling-product operators for the XYZ+h model and extended doubling-product operators for the NNN-XYZ model) are zero.
At the same time, we also compute explicit expressions of the remaining coefficients of $\bsA^k$.
In particular, we show that if one of the remaining coefficients is zero, then all the remaining coefficients must be zero.
\item We examine the conditions for $r_{\bsB^k}=0$ for all $\bsB^k$, and show that one of the remaining coefficient of $\bsA^k$ is zero.
\end{enumerate}
The latter step is highly model-dependent, and we need elaborated constructions for the XYZ+h model and the NNN-XYZ model separately.
In contrast, the former steps for these models are almost the same.

\subsection{Symbols and terms (1)}\lb{s:symbol1}


In this paper, when we align Pauli operators as $XY$, this symbol means an operator where $X$ acts on a site and $Y$ acts on the next site (i.e., $X_i\otimes Y_{i+1}$).
If we intend to express a product of Pauli operators on the same site, we use a dot symbol $\cdot$ as $X\cdot Y$.
For completeness, we summarize the rule of the product of Pauli matrices below:
\balign{
X\cdot X=Y\cdot Y=Z\cdot Z=&I, \\
X\cdot Y=-Y\cdot X=&iZ, \\
Y\cdot Z=-Z\cdot Y=&iX, \\
Z\cdot X=-X\cdot Z=&iY.
}
This rule leads to an expression of a commutator of two different Pauli matrices $A,B \in \{ X,Y,Z\}$ ($A\neq B$) as
\eqa{
[A, B]=2A\cdot B.
}{commutator-product}

For our later use, it is convenient to define {\it divestment} of a phase factor and {\it signless product} of Pauli matrices.
We {\it divest} a phase factor of Pauli matrices as
\eq{
|aX|=|a|X, \hspace{10pt}
|aY|=|a|Y, \hspace{10pt}
|aZ|=|a|Z
}
with $a\in \bbC$, where the symbol $|\cdot |$ denote divestment.
The signless product of Pauli matrices is a product with divestment, which is written as
\balign{
|X\cdot Y|=|Y\cdot X|=&Z, \\
|Y\cdot Z|=|Z\cdot Y|=&X, \\
|Z\cdot X|=|X\cdot Z|=&Y.
}
To recover the lost phase factor, we introduce the sign factor $\sigma(A, B)=\pm 1$ for $A,B\in \{ X,Y,Z\}$ ($A\neq B$)  so that
\eq{
A\cdot B=i\sigma(A, B)|A\cdot B|.
}
A concrete expression is
\balign{
\sigma(X,Y)=\sigma(Y,Z)=\sigma(Z,X)=&1, \\
\sigma(Y,X)=\sigma(Z,Y)=\sigma(X,Z)=&-1.
}

For a product of $l$ operators $\bsA=A_i^1A_{i+1}^2\cdots A_{i+l-1}^l$, the divestment is defined as
\eq{
|\bsA|=|A_i^1||A_{i+1}^2|\cdots |A_{i+l-1}^l|.
}
If $A^i\in \{X,Y,Z\}$ and $A^i\neq A^{i+1}$ are satisfied for all $i$, its sign factor is defined recursively as
\balign{
\sigma(A^1,A^2,\ldots, A^l)=\sigma(A^1,A^2)\sigma(A^2,A^3,\ldots , A^l)=&\sigma(A^1,A^2)\sigma(A^2,A^3)\sigma(A^3,\ldots , A^l) \nt \\
=&\sigma(A^1,A^2)\sigma(A^2,A^3)\cdots \sigma(A^{l-1},A^l). \lb{def-sigma-gen}
}
For example, we have $\sigma(X,Y,X,Z)=\sigma(X,Y)\sigma(Y,X)\sigma(X,Z)=1$.

\bigskip

We shall introduce some further symbols and terms to describe commutators.
When a commutation relation $[\bsA, \bsC]=c\bsD$ holds with a number coefficient $c$, we say that the operator $\bsD$ is {\it generated} by the commutator of $\bsA$ and $\bsC$.
In our proof, we examine commutators generating a given operator and derive a relation of coefficients of operators.

Consider the XYZ+h model and a candidate of conserved quantity $Q$ with $k=4$.
A 5-support operator $\sum_i X_iY_{i+1}X_{i+2}Y_{i+3}X_{i+4}$ in $\qh$, for example, is generated by the following two commutators:
\balign{
-\i [X_iY_{i+1}X_{i+2}Z_{i+3}, Y_{i+3}Y_{i+4}]&=-2X_iY_{i+1}X_{i+2}X_{i+3}Y_{i+4}, \\
-\i [Z_{i+1}X_{i+2}X_{i+3}Y_{i+4}, X_iX_{i+1}]&=2X_iY_{i+1}X_{i+2}X_{i+3}Y_{i+4},
}
where we dropped $\sum_i$, the summation over $i$, for visibility.
In case without confusion, we also drop subscripts for brevity.
We visualize these two commutation relations similarly to the column addition as follows:
\eq{
\begin{array}{rccccc}
&X&Y&X&Z& \\
&&&&Y&Y \\ \hline
-2&X&Y&X&X&Y
\end{array}
\ \ \ \ \
\begin{array}{rccccc}
&&Z&X&X&Y \\
&X&X&&& \\ \hline
2&X&Y&X&X&Y
\end{array}. \nt 
}
Here, two arguments of the commutator are written above the horizontal line, and the result of the commutator(including the imaginary number $\i$) is written below the horizontal line.
The horizontal positions in this visualization represent the spatial positions of spin operators.

The operator $XYXXY$ in $[Q,H]$ is generated only by the above two commutators, from which we say that $XYXZ$ and $ZXXY$ form a {\it pair}.
Then, the condition $r_{XYXXY}=0$ implies the following relation
\eq{
-q_{XYXZ}+q_{ZXXY}=0.
}

\bigskip

We next introduce a useful symbol $\a_i :=A_iA_{i+1}$, which we call as {\it doubling-product}.
The Hamiltonian of the XYZ+h model contains three doubling-products, $\x=XX$, $\y=YY$, and $\z=ZZ$.
When we align several doubling-products, we promise that a neighboring doubling-product has its support with a single-site shift.
For example, we can express
\eqa{
\x\y\x _i=|(X_iX_{i+1})(Y_{i+1}Y_{i+2})(X_{i+2}X_{i+3})| =X_i|X_{i+1}\cdot Y_{i+1}||Y_{i+2}\cdot X_{i+2}||X_{i+3} =X_i Z_{i+1} Z_{i+2}X_{i+3}.
}{doubling-product-def}
We require that the same doubling-products cannot be neighboring (e.g., $\x\x\z$ is not allowed).
We call operators expressed in the above form as {\it doubling-product operators}.
In general, a doubling-product operator $\overline{\bsA}$ with $l$ doubling-products $\overline{A^1}, \overline{A^2},\ldots ,\overline{A^l}$  is computed as
\eq{
\overline{\bsA}_i=A_i^1|A_{i+1}^1\cdot A_{i+1}^2||A_{i+2}^2\cdot A_{i+2}^3|\cdots |A_{i+l-1}^{l-1}\cdot A_{i+l-1}^l| A_{i+l}^l.
}
Using the sign factor $\sigma$ defined in \eref{def-sigma-gen}, the phase factor of $\overline{\bsA}_i$ can be recovered as
\eq{
(A_i^1A_{i+1}^1)\cdot (A_{i+1}^2A_{i+2}^2)\cdots (A_{i+l-1}^l A_{i+l}^l)=(\i)^l \sigma(A^1,A^2,\ldots, A^l) \overline{\bsA}_i.
}

To highlight the power of the expression with doubling-products, we write a doubling-product operator $\overline{ABC\cdots D}$ as
\eqa{
\ba{cccccc}{
A&A&&&& \\
&B&B&&& \\
&&C&C&& \\
&&&\ddots&& \\
&&&&D&D \\ \hline \hline
},
}{ladder}
which we call a {\it column expression} of  $\overline{ABC\cdots D}$.
Here, the double horizontal line means the multiplication of all the operators with divestment (removing the phase factor), which we use for both doubling-product and non-doubling-product operators.
Keep in mind not to confuse a single horizontal line, which represents a commutation relation.
Single and double horizontal lines are frequently used at the same time as
\eqa{
\ba{ccccccc}{
A&A&&&&& \\
&B&B&&&& \\
&&C&C&&& \\
&&&\ddots&&& \\
&&&&D&D& \\ \hline \hline
&&&&&E&E \\ \hline
},
}{ex-ABCD-E}
which represents the commutator $[\overline{ABC\cdots D}, \overline{E}]$.
Here we abbreviated the last row (the resulting operator of these commutators) for brevity.
The column expression can be easily extended to operators which are not doubling-product operators.

Similarly to doubling-product operators, we define the sign of a commutator $[A,B]$ including $-\i$ denoted by $s([A,B])$ as
\eq{
-\i [A,B]=2s([A,B])|A\cdot B|.
}
For example, we have $s([\z_i, \y_{i+1}])=-1$, which follows from $-\i[\z_i, \y_{i+1}]=-2 ZXY_i$ and $|\z_i \cdot \y_{i+1}|=ZXY_i$.
If both $A$ and $B$ are single Pauli matrices on the same site, we have $s([A,B])=\sigma(A,B)$.
We also express the argument of $s$ in the form of the horizontal line as
\eq{
s([XZY_i, ZZ_{i+2}])=
s\( \ba{cccc}{
X&Z&Y& \\
&&Z&Z \\ \hline
} \)
=1,
}
where $XZY_i$ and $ZZ_{i+2}$ are the abbreviations of $X_iZ_{i+1}Y_{i+2}$ and $Z_{i+2}Z_{i+3}$, respectively.
We promise that if an operator has a single subscript, the subscript represents the leftmost site on which this operator acts.
Using the sign of a commutator, the commutator \eref{ex-ABCD-E} is expressed as
\eq{
[\overline{ABC\cdots D}, \overline{E}]=2\i s([D,E])\overline{ABC\cdots DE}.
}

\section{XYZ model with Z magnetic field: a review}\lb{s:XYZh}

\subsection{Commutators generating $k+1$-support operators}\lb{s:XYZh-step1}

We prove the absence of nontrivial local conserved quantity along the idea presented in \sref{proof-idea}.
Suppose that a $k$-support operator $Q$ is a conserved quantity.
The conservation of $Q$ implies that all $r_{\bsB^l}$ in \eref{QH} is zero; $r_{\bsB^l}=0$.
Using this fact, we derive many relations on the coefficients $q_{\bsA}$ in \eref{Qform} by comparing both sides of \eref{QH} and show that $q_{\bsA ^k}=0$ for all $\bsA^k$.

A commutator of a $k$-support operator and the Hamiltonian can generate an at most $k+1$-support operator.
A $k+1$-support operator in $\qh$ is generated by a commutator such that the interaction term ($XX$, $YY$, and $ZZ$) acts on the left end or the right end of a $k$-support operator as
\eq{
\ba{ccccc}{
A&B&\cdots&C& \\
&&&X&X \\ \hline
A&B&\cdots&D&X
}
, \ \
\ba{ccccc}{
&E&\cdots&F&G \\
Y &Y&&& \\ \hline
Y &H&\cdots&F&G
}.
}
In the first step of our proof (step a in \sref{proof-idea}), we consider the case that the commutator generates $k+1$-support operators.
The analysis on commutators generating $k+1$-support operators leads to the following consequence:

\blm{
Let $Q$ be a $k$-support conserved quantity expanded as \eref{Qform}.
Then, its coefficients of a $k$-support operator $\bsA\in \calP^k$ should be expressed as
\eqa{
q_{\bsA}=\cX^k \cdot \sigma(B^1,B^2,\ldots , B^{k-1}) \prod_{i=1}^{k-1} J^{B^i}
}{doubling-coefficient}
with a common constant $\cX^k $ if $\bsA$ is a doubling-product operator written as $\bsA=\prod_{i=1}^{k-1} \overline{B}^i$, and $q_{\bsA}=0$ otherwise.
}


Let us prove Lemma 1.
We first treat the case that $\bsA$ is a doubling-product operator and confirm \eref{doubling-coefficient}.
To this end, we consider two commutators generating the same $k+1$-support operator in the column expression as
\eqa{
\ba{ccccccc}{
A&A&&&&& \\
&B&B&&&& \\
&&C&C&&& \\
&&&\ddots&&& \\
&&&&D&D& \\ \hline \hline
&&&&&E&E \\ \hline
}
, \ \ \ \
\ba{ccccccc}{
&B&B&&&& \\
&&C&C&&& \\
&&&\ddots&&& \\
&&&&D&D& \\
&&&&&E&E \\ \hline \hline
A&A&&&&& \\ \hline
}
}{doubling-pair}
with $A,B,C,\ldots ,D,E\in \{ X,Y,Z\}$.
These two diagrams represent commutators $[\overline{ABC\cdots D}_i, \overline{E}_{i+k-1}]$ and $[\overline{BC\cdots DE}_{i+1}, \a_i]$, respectively.
Since the $k+1$-support operator $A|B\cdot C|\cdots |D\cdot E|E$ is generated only by these two commutators, these two doubling-product operators $\overline{ABC\cdots D}$ and $\overline{BC\cdots DE}$ form a pair, and we have
\eqa{
J^E q_{\overline{ABC\cdots D}}[\overline{ABC\cdots D}_i, \overline{E}_{i+k-1}]-J^A q_{\overline{BC\cdots DE}}[\overline{BC\cdots DE}_{i+1}, \a_i]=0.
}{XYZ-step1-mid1}

The signs of these two commutators in \eref{doubling-pair} are computed as
\balign{
s\( \ba{ccccccc}{
A&A&&&&& \\
&B&B&&&& \\
&&C&C&&& \\
&&&\ddots&&& \\
&&&&D&D& \\ \hline \hline
&&&&&E&E \\ \hline
} \)
&=s([D,E])=\frac{\sigma(\overline{ABC\cdots DE})}{\sigma(\overline{ABC\cdots D})}, \lb{commute-plus} \\
s\( \ba{ccccccc}{
&B&B&&&& \\
&&C&C&&& \\
&&&\ddots&&& \\
&&&&D&D& \\
&&&&&E&E \\ \hline \hline
A&A&&&&& \\ \hline
} \)
&=s([B,A])=-\frac{\sigma(\overline{ABC\cdots DE})}{\sigma(\overline{BC\cdots DE})}, \lb{commute-minus}
}
respectively.
The latter has the minus sign because the order of $\a$ and $\b$ in multiplication (or commutation) is converted compared to the order of products in the definition of the column expression.
In summary, two commutators in \eref{doubling-pair} imply a relation of coefficients:
\eqa{
\frac{q_{\overline{ABC\cdots D}}}{\sigma(A,B,C,\ldots, D)}J^E = \frac{q_{\overline{BC\cdots DE}}}{\sigma(B,C,\ldots, D,E)}J^A
}{Lemma1-first-mid1}
for any sequence $ABC\cdots DE$, which is equivalent to
\eqa{
\frac{q_{\overline{ABC\cdots D}}}{\sigma(A,B,C,\ldots, D)J^AJ^BJ^C\cdots J^D} = \frac{q_{\overline{BC\cdots DE}}}{\sigma(B,C,\ldots, D,E)J^BJ^C\cdots J^DJ^E}.
}{Lemma1-first-mid2}
Fix $\overline{ABC\cdots D}$ as a reference doubling-product operator with length $k$, whose coefficient is written in the form of Lemma 1 as
\eq{
q_{\overline{ABC\cdots D}}=\cX^k \cdot \sigma(\overline{ABC\cdots D}) J^AJ^BJ^C\cdots J^D
}
with some $\cX^k$.
Then, using \eref{Lemma1-first-mid2}, the coefficient of $\overline{BC\cdots DE}$ is also shown to be written in the form of Lemma 1 as
\eq{
q_{\overline{BC\cdots DE}}=\cX^k \cdot \sigma(\overline{BC\cdots DE}) J^BJ^C\cdots J^DJ^E.
}
Since any pair of doubling-product operators is connected as
\balign{
\overline{B^1B^2B^3\cdots B^{k-1}}\lr& \overline{B^2B^3\cdots B^{k-1}D}\lr \overline{B^3\cdots B^{k-1}DC^1}\lr \cdots  \nt \\
\cdots \lr& \overline{B^{k-1}DC^1\cdots C^{k-3}}
\lr \overline{DC^1\cdots C^{k-3}C^{k-2}} \lr \overline{C^1\cdots C^{k-3}C^{k-2}C^{k-1}},
}
repeated applications of \eref{Lemma1-first-mid2} implies the first part of Lemma 1.

\bigskip

We next clarify what happens if an operator $\bsA$ is not a doubling-product operator, which is the second part of Lemma 1.
In a non-doubling-product operator $\bsA$, one of the following three happens.
\be{
\item One of $A^2\ldots , A^{k-1}$ is an identity operator $I$.
\item There exists $2\leq n\leq k-1$ satisfying $\abs{\prod_{i=1}^{n-1}A^i}=\abs{A^1\cdot A^2\cdots A^{n-1}}=A^n$.
\item $\prod_{i=1}^{k-1}A^i=\abs{A^1\cdot A^2\cdots A^{k-1}}\neq A^k$ holds.
}
To confirm the above fact, it suffices to demonstrate that an operator $\bsA$ without the above three conditions is indeed a doubling-product operator.
Since $\abs{\prod_{i=1}^{n-1}A^i}\neq A^n$ holds for $2\leq n\leq k-1$ (negation of condition 2), we notice that $B^n=\abs{\prod_{i=1}^{n}A^i}=\abs{\prod_{i=1}^{n-1}A^i \cdot A^n}$ for $2\leq n\leq k-1$ are Pauli operators (not the identity operator).
By construction, $B^n B^{n+1}=\abs{\prod_{i=1}^{n-1}A^i\prod_{i=1}^{n}A^i}=A^n$ is satisfied, and the negation of condition 1 suggests that $B^n B^{n+1}=A^n$ is not the identity operator.
In addition, $\prod_{i=1}^{k-1}A^i=A^k$ (the negation of condition 3) implies $B^{k-1}=A^k$.
Hence, $\bsA$ is a doubling-product operator $\bsA=\overline{A^1B^2B^3\cdots B^{k-1}}$.

Below we shall explain why the coefficient becomes zero in these three cases.
We first demonstrate the idea by simple examples and then provide a general proof.
In case 1, the column expression of an operator, e.g., $XYXZIZXY$, is written as
\eq{
XYXZIZXY=\ba{cccccccc}{
X&X&&&&&& \\
&Z&Z&&&&& \\
&&Y&Y&&&& \\
&&&X&I&&& \\
&&&&&Z&Z& \\
&&&&&&Y&Y \\
\hline \hline
}.
}
This operator forms a pair with $YZYXZIZZ$, which is seen in generating $YZYXZIZXY$:
\eq{
\ba{ccccccccc}{
&X&X&&&&&& \\
&&Z&Z&&&&& \\
&&&Y&Y&&&& \\
&&&&X&I&&&\\
&&&&&&Z&Z& \\
&&&&&&&Y&Y \\
\hline \hline
Y&Y&&&&&&& \\
\hline
},
\hspace{10pt}
\ba{ccccccccc}{
Y&Y&&&&&&& \\
&X&X&&&&&& \\
&&Z&Z&&&&& \\
&&&Y&Y&&&& \\
&&&&X&I&&&\\
&&&&&&Z&Z& \\
\hline \hline
&&&&&&&Y&Y \\
\hline
}.
}
These two commutators suggest
\eq{
J^Y q_{XYXZIZXY}-J^Yq_{YZYXZIZZ}=0.
}
However, $YZYXZIZZ$ cannot form a pair in generating $XZZYXZIZZ$
\eq{
\ba{cccccccccc}{
&&Y&Y&&&&&& \\
&&&X&X&&&&& \\
&&&&Z&Z&&&& \\
&&&&&Y&Y&&& \\
&&&&&&X&I&& \\
&&&&&&&&Z&Z \\
\hline \hline
&X&X&&&&&&& \\
\hline
-2&X&Z&Z&Y&X&Z&I&Z&Z
}
}
because the right end is $ZZ$ and no commutator of 8-support operator and $XX$, $YY$, $ZZ$ can result in this form of operators:
\eq{
\ba{cccccccccc}{
&?&?&?&?&?&?&?&?& \\
&&&&&&&&Z&Z \\
\hline
&X&Z&Z&Y&X&Z&I&Z&Z
}.
}
Since the coefficient of $XZZYXZIZZ$ in $\qh$ is zero, our observation directly means
\eq{
q_{YZYXZIZZ}=0,
}
which leads to
\eq{
q_{XYXZIZXY}=0.
}

Cases 2 and 3 can be treated similarly to case 1.
An example of case 2 in the column expression is
\eq{
XYZXZXXY=\ba{cccccccc}{
X&X&&&&&& \\
&Z&Z&&&&& \\
&&&X&X&&& \\
&&&&Y&Y&& \\
&&&&&Z&Z& \\
&&&&&&Y&Y \\
\hline \hline
},
}
which forms a pair with $ZZXZXXXZ$ as
\eq{
\ba{ccccccccc}{
X&X&&&&&&& \\
&Z&Z&&&&&& \\
&&&X&X&&&& \\
&&&&Y&Y&&& \\
&&&&&Z&Z&& \\
&&&&&&Y&Y& \\
\hline \hline
&&&&&&&Z&Z \\
\hline
},
\hspace{10pt}
\ba{ccccccccc}{
&Z&Z&&&&&& \\
&&&X&X&&&& \\
&&&&Y&Y&&& \\
&&&&&Z&Z&& \\
&&&&&&Y&Y& \\
&&&&&&&Z&Z \\
\hline \hline
X&X&&&&&&& \\
\hline
}.
}
However, $ZZXZXXXZ$ does not form a pair in generating $ZZXZXXXXY$
\eq{
\ba{cccccccccc}{
&Z&Z&&&&&&& \\
&&&X&X&&&&&\\
&&&&Y&Y&&&& \\
&&&&&Z&Z&&& \\
&&&&&&Y&Y&& \\
&&&&&&&Z&Z& \\
\hline \hline
&&&&&&&&Y&Y \\
\hline
-2&Z&Z&X&Z&X&X&X&X&Y
},
}
because the left end is $ZZ$, and no commutator of 8-support operator and $XX$, $YY$, $ZZ$ can result in this form of operators:
\eq{
\ba{cccccccccc}{
&&?&?&?&?&?&?&?&? \\
&Z&Z&&&&&&& \\
\hline
-2&Z&Z&X&Z&X&X&X&X&Y
}.
}
This directly implies
\eq{
q_{ZZXZXXXZ}=0,
}
which leads to
\eq{
q_{XYZXZXXY}=0.
}

An example of case 3 in the column expression is
\eq{
XYXZZXXZ=\ba{cccccccc}{
X&X&&&&&& \\
&Z&Z&&&&& \\
&&Y&Y&&&& \\
&&&X&X&&& \\
&&&&Y&Y&& \\
&&&&&Z&Z& \\
&&&&&&Y&Z \\
\hline \hline
},
}
where the last line is not $YY$ but $YZ$.
Here, $XYXZZXXZ$ forms a pair with $ZXZZXXYX$ in generating $XYXZZXXYX$:
\eq{
\ba{ccccccccc}{
X&X&&&&&&& \\
&Z&Z&&&&&& \\
&&Y&Y&&&&& \\
&&&X&X&&&& \\
&&&&Y&Y&&& \\
&&&&&Z&Z&& \\
&&&&&&Y&Z& \\
\hline \hline
&&&&&&&X&X \\
\hline
},
\hspace{10pt}
\ba{ccccccccc}{
&Z&Z&&&&&& \\
&&Y&Y&&&&& \\
&&&X&X&&&& \\
&&&&Y&Y&&& \\
&&&&&Z&Z&& \\
&&&&&&Y&Z& \\
&&&&&&&X&X \\
\hline \hline
X&X&&&&&&& \\
\hline
}.
}
As a consequence, the coefficient $q_{XYXZZXXZ}$ is connected to $q_{ZXZZXXYX}$ with multiplying some constant.

Similarly to previous arguments, we can remove doubling-products from the left and add doubling-products to the right repeatedly, which results in the connection to
\eq{
\ba{cccccccc}{
X&X&&&&&& \\
&Z&Z&&&&& \\
&&Y&Y&&&& \\
&&&X&X&&& \\
&&&&Y&Y&& \\
&&&&&Z&Z& \\
&&&&&&Y&Z \\
\hline \hline
}\lr
\ba{cccccccc}{
Z&Z&&&&&& \\
&Y&Y&&&&& \\
&&X&X&&&& \\
&&&Y&Y&&& \\
&&&&Z&Z&& \\
&&&&&Y&Z& \\
&&&&&&X&X \\
\hline \hline
}
\lr\cdots \lr
\ba{cccccccc}{
Y&Z&&&&&& \\
&X&X&&&&& \\
&&Y&Y&&&& \\
&&&X&X&&& \\
&&&&Y&Y&& \\
&&&&&X&X& \\
&&&&&&Y&Y \\
\hline \hline
}.
}
However, the last operator $YYZZZZZY$ does not form a pair in generating $YYZZZZZZX$ because the left end is $YY$.
Following a similar argument to previous ones, we conclude that the coefficient of the initial operator, $q_{XYXZZXXZ}$, is zero.

\bigskip

Now we shall formulate this argument in a general form.
Our starting point is the fact that if $A^1=A^2$ or $A^2=I$, then $q_{\bsA}=0$.
This fact holds because $A^1A^2\cdots A^{k-1}|A^k\cdot B|B$ with $B\neq A^k$ is generated only by a commutator $[\bsA_i, \b_{i+k-1}]$.

To state our key observation for our proof, we construct a sequence of Pauli operators $\{ B^n\}$ with $B^n\neq B^{n+1}$ and  $B^1\neq A^k$.
We denote $\abs{\prod_{i=1}^n A^i}=K^n$ and $C^n=\abs{B^{n-1} B^n}$ with regarding $B^0=A^k$ for convenience.
We observe that if $A^1\neq A^2$ and $A^2\neq I$, then by considering commutators generating $A^1A^2A^3\cdots A^{k-1}C^1B^1$, we have a linear relation between the coefficient of $\bsA=A^1A^2A^3\cdots A^{k-1}A^k$ and $K^2A^3\cdots A^{k-1}C^1B^1=|A^1\cdot A^2|A^3\cdots A^{k-1}|A^k\cdot B^1|B^1$ as
\eq{
J^{B^1} s([A^k, B^1])q_{\bsA}=J^{A^1} s([A^2, A^1]) q_{K^2A^3\cdots A^{k-1}C^1B^1}.
}
In a similar manner, if $K^2\neq A^3$ and $A^3\neq I$, we have
\eq{
J^{B^2} s([B^1, B^2]) q_{K^2A^3\cdots A^{k-1}C^1B^1}=J^{K^2} s([K^3,K^2])q_{K^3A^4A^5\cdots A^{k-1}C^1C^2B^2}.
}
In general, if $K^n\neq A^{n+1}$ and $A^{n+1}\neq I$, we have
\eq{
J^{B^n} s([B^{n-1}, B^n]) q_{K^nA^{n+1}\cdots A^{k-1}C^1C^2\cdots C^{n-1}B^{n-1}}=J^{K^n} s([K^{n+1}, K^n])q_{K^{n+1}A^{n+2}\cdots A^{k-1}C^1C^2\cdots C^{n}B^{n}},
}
and consequently $q_{\bsA}$ and $q_{K^{n+1}A^{n+2}\cdots A^{k-1}C^1C^2\cdots C^{n}B^{n}}$ are linearly connected through these relations.

We shall show that if $\bsA$ is a non-doubling product operator, then $q_{K^{n+1}A^{n+2}\cdots A^{k-1}C^1C^2\cdots C^{n}B^{n}}=0$ for some $n$, which leads to the desired result $q_{\bsA}=0$.
We first consider cases 1 and 2.
By assumption, there exists $n$ such that $K^{n+1}=A^{n+2}$ or $A^{n+2}=I$.
Then, $q_{K^{n+1}A^{n+2}\cdots A^{k-1}C^1C^2\cdots C^{n}B^{n}}=0$ holds for this $n$ because $K^{n+1}A^{n+2}\cdots A^{k-1}C^1C^2\cdots C^{n}C^{n+1}B^{n+1}$ is generated only by a commutator $[K^{n+1}A^{n+2}\cdots A^{k-1}C^1C^2\cdots C^{n}B^{n}_i, \overline{B^{n+1}}_{i+k-1}]$.

We next consider case 3.
Since $K^{k-1}\neq A^k$, we can set $B^1$ such that $C^1=\abs{A^k\cdot B^1}=K^{k-1}$.
The aforementioned linear relation is elongated to the case of $n=k-2$, where the coefficient $q_{K^{k-1}C^1C^2\cdots C^{k-2}B^{k-2}}$ is in consideration.
Then, $q_{K^{k-1}C^1C^2\cdots C^{k-2}B^{k-2}}=0$ holds because $q_{K^{k-1}C^1C^2\cdots C^{k-2}C^{k-1}B^{k-1}}$ is generated only by a commutator $[K^{k-1}C^1C^2\cdots C^{k-2}B^{k-2}, \overline{B^{k-1}}_{i+k-1}]$ (Notice $K^{k-1}=C^1$).
This completes the proof of Lemma 1.

\bigskip

In summary, we used the fact that (1) a connection of a pair with a commutator can be regarded as removing and adding a doubling-product at the left or right end, and (2) if two leftmost sites or two rightmost sites of a $k$-support operator are the same Pauli operator ($XX$, $YY$, or $ZZ$), then it cannot form a pair and thus has zero coefficient.
This idea is quite general and is also used in treating the NNN-XYZ model in \sref{NNN}.

\subsection{Demonstration with $k=3$; an example}\lb{s:k=3demo}
Before going to our second step (step b in \sref{proof-idea}) for general $k$, we here present the idea of step b in the simplest case, the case of $k=3$.

In step b for $k=3$, we treat 3-support operators in $[Q,H]$.
First, $YZY$ is generated by the following four commutators;
\eq{
\begin{array}{rccc}
&Y&Z&X \\
&&&Z \\ \hline
-2&Y&Z&Y
\end{array}
\ \ \ \ \
\begin{array}{rccc}
&X&Z&Y \\
&Z&& \\ \hline
-2&Y&Z&Y
\end{array}
\ \ \ \ \
\begin{array}{rccc}
&Y&X& \\
&&Y&Y \\ \hline
2&Y&Z&Y
\end{array}
\ \ \ \ \
\begin{array}{rccc}
&&X&Y \\
&Y&Y& \\ \hline
2&Y&Z&Y.
\end{array} \nt
}
We note that $YZX=\y\x$ and $XZY=\x\y$ are doubling-product operators.
Since $YZY$ is generated only by these four commutators, we have
\eqa{
h(q_{YZX}+q_{XZY})-J_Y(q_{YX}+q_{XY})=0.
}{3-YZY}

Similarly to this, we consider $YYZ$
\eq{
\begin{array}{rccc}
&X&Y&Z \\
&Z&& \\ \hline
-2&Y&Y&Z
\end{array}
\ \ \ \
\begin{array}{rccc}
&Y&X&Z \\
&&Z& \\ \hline
-2&Y&Y&Z
\end{array}
\ \ \ \
\begin{array}{rccc}
&Y&X& \\
&&Z&Z \\ \hline
-2&Y&Y&Z
\end{array}  \nt
}
and $XXZ$
\eq{
\begin{array}{rccc}
&X&Y&Z \\
&&Z& \\ \hline
2&X&X&Z
\end{array}
\ \ \ \
\begin{array}{rccc}
&Y&X&Z \\
&Z&& \\ \hline
2&X&X&Z
\end{array}
\ \ \ \
\begin{array}{rccc}
&X&Y& \\
&&Z&Z \\ \hline
2&X&X&Z,
\end{array}  \nt
}
both of which are generated only by three commutators.
Again, we note that $XYZ=\x\z$ and $YXZ=\y\z$ are doubling-product operators.
These two sets of commutators imply
\balign{
h(q_{XYZ}+q_{YXZ})+J_Zq_{YX}=&0, \lb{3-YYZ} \\
h(q_{XYZ}+q_{YXZ})+J_Zq_{XY}=&0. \lb{3-XXZ}
}
Combining Eqs.~\eqref{3-YZY}, \eqref{3-YYZ}, and \eqref{3-XXZ} to erase $q_{YX}$ and $q_{XY}$, and inserting \eref{doubling-coefficient} shown in lemma 1, we obtain
\eq{
h\( 1-\frac{J_Y}{J_X}\) \cX^3 J^XJ^Z=0,
}
which directly implies the desired result $\cX^3=0$ as long as $h\neq 0$ and $J_X\neq J_Y$.

\subsection{Symbols and terms (2)}\lb{s:symbol2}
We are ready to prove Theorem 1 for general $k$ according to step b (in \sref{proof-idea}) by analyzing $k$-support operators in $\qh$.
We again use the expression \eqref{Qform}.
We have shown in Lemma 1 that $q_{\bsA}=0$ for all non-doubling-product operators $\bsA\in \calP^k$ and that all the remaining $q_{\bsA}$ with $\bsA\in \calP^k$ is linearly connected.
Thus, it suffices to demonstrate that a doubling-product operator $\bsA\in \calP^k$ has zero coefficient, $q_{\bsA}=0$.
Toward this end, we introduce further symbols.

We shall introduce symbols by taking some examples.
In the case of $k=5$, $ZXZXZ$ is generated by the following four commutators:
\eqa{
\ba{rccccc}{
&Z&X&Z&Y&Z \\
&&&&Z& \\ \hline
2&Z&X&Z&X&Z,
}
\ \
\ba{rccccc}{
&Z&Y&Z&X&Z \\
&&Z&&& \\ \hline
2&Z&X&Z&X&Z,
}
\ \
\ba{rccccc}{
&Z&X&Z&Y& \\
&&&&Z&Z \\ \hline
2&Z&X&Z&X&Z,
}
\ \
\ba{rccccc}{
&&Y&Z&X&Z \\
&Z&Z&&& \\ \hline
2&Z&X&Z&X&Z.
}
}{ZXZXZ}
Using the column expression, these commutators read
\eqa{
\ba{ccccc}{
Z&Z&&& \\
&Y&Y&& \\
&&X&X& \\
&&&Z&Z \\ \hline \hline
&&&Z& \\ \hline
},
\ \
\ba{ccccc}{
Z&Z&&& \\
&X&X&& \\
&&Y&Y& \\
&&&Z&Z \\ \hline \hline
&Z&&& \\ \hline
},
\ \
\ba{ccccc}{
Z&Z&&& \\
&Y&Y&& \\
&&X&X& \\
&&&Z& \\ \hline \hline
&&&Z&Z \\ \hline
},
\ \
\ba{ccccc}{
&X&X&& \\
&&Y&Y& \\
&&&Z&Z \\
&Z&&& \\ \hline \hline
Z&Z&&& \\ \hline
}.
}{ZXZXZ-2}
A key fact behind the first two commutators is the relation
\eq{
\ba{ccccc}{
Z&Z&&& \\
&Y&Y&& \\
&&X&X& \\
&&&Z&Z \\
&&&Z& \\ \hline \hline
}=
\ba{ccccc}{
Z&Z&&& \\
&X&X&& \\
&&Y&Y& \\
&&&Z&Z \\
&Z&&& \\ \hline \hline
},
}
where one should recall \eref{commutator-product} implying that a signless product (double line) and a commutation relation (single line) result in the same operator.
This equality suggests that we can switch the role of $\x$ and $\y$ in the alternation of these two by moving a single $Z$ from one end to the other.
As will be demonstrated at the beginning of \sref{XYZh-step2}, we shall extend this idea to general strings.

Now we introduce some symbols which help to describe commutators as above.
First, we introduce a symbol ``$\hz$" representing a commutator with $Z$ at which the left and right doubling-products have an overlap.
For example, $\x\y\hz\z$ represents the commutator $[X_iZ_{i+1}X_{i+2}Z_{i+3}, Z_{i+2}]$, since $\x\y\z_i=\abs{(X_iX_{i+1})\cdot (Y_{i+1}Y_{i+2})\cdot (Z_{i+2}Z_{i+3})}$ and the overlap of $\y$ and $\z$ is at site $i+2$.
Using this symbol, the first two commutators in \eqref{ZXZXZ} (see also \eref{ZXZXZ-2}) are expressed simply as
\eq{
\z\y\x\hz\z , \ \ \ \ \z\hz\x\y\z . \nt
}

Next, we introduce a symbol ``$\bz$", which represents the multiplication of $Z$ at this position in the column expression.
Here, we employ the signless product in the multiplication.
For example, $\x\bz\z\y\x$ means
\eq{
\x\bz\z\y\x=
\ba{ccccc}{
X&X&&& \\
&Z&Z&& \\
&&Y&Y& \\
&&&X&X \\
&Z&&& \\ \hline \hline
}
=XXXZX.
}
We note that the expression with $\bz$ is not unique (e.g., $ZXZY=\z\y\x\bz=\z\bz\x\y$).

We further introduce two symbols ``$\rplus$" and ``$\lplus$", which mean that commutators act at the rightmost and leftmost sites, respectively.
Examples are $\x\y \rplus \x =[\x\y_i, \x_{i+2}]$ and $\z \lplus YYYZ=[\z_i, Y_{i+1}Y_{i+2}Y_{i+3}Z_{i+4}]$.
Then, the latter two commutators in \eqref{ZXZXZ} (see also \eref{ZXZXZ-2}) are expressed as
\eq{
\z\y\x\bz \rplus \z , \ \ \ \ \  \z \lplus \bz\x\y\z . \nt
}

We finally introduce symbols which represents alternating $\x$ and $\y$ defined as
\balign{
L^{2n}&:=\underbrace{\overline{YX\cdots YX}}_{n \text{ copies of }\y\x}, \\
L^{2n+1}&:=\overline{X\underbrace{YX\cdots YX}_{n \text{ copies of }\y\x}}, \lb{def-L2n+1} \\
R^{2n}&:=\underbrace{\overline{XY\cdots XY}}_{n \text{ copies of }\x\y}, \\
R^{2n+1}&:=\overline{\underbrace{XY\cdots XY}_{n \text{ copies of }\x\y}}X. \lb{def-R2n+1}
}
Correspondingly, when we consider the sign $\sigma$ of a string with these symbols, we promise that $L$ and $R$ mean $Y,X,Y,X\ldots$ and $X,Y,X,\ldots$, respectively.
For example, $\sigma(L^4, Z, Y, R^3)$ means $\sigma(Y,X,Y,X,Z,Y,X,Y,X)$.

\subsection{Commutators generating $k$-support operators}\lb{s:XYZh-step2}

In general, as seen in the previous subsection in \eref{ZXZXZ}, a single $k$-support operator in \qh is generated by four commutators; two are of a $k$-support operator in $Q$ and a magnetic field (1-support operator) in $H$, and the other two are of a $k-1$-support operator in $Q$ and the exchange interaction (2-support operator) in $H$.
Concretely, the following four commutators
\balign{
&\overline{AB\cdots  Z} \hz \overline{X  Y  X  Y \cdots  X  Y  Z \cdots CD} , \\
&\overline{AB\cdots  Z  Y X  Y  X  \cdots   Y X} \hz \overline{Z \cdots CD} ,  \\
&\overline{AB\cdots  Z} \bz \overline{X  Y  X  Y \cdots  X  Y  Z \cdots C} \rplus \d ,  \\
&\a \lplus \overline{B\cdots  Z} \bz \overline{X  Y  X  Y \cdots  X  Y  Z \cdots CD}
}
generate the same operator.
Note that $\overline{\cdots XYZ \cdots}$ might be $\overline{\cdots YXZ\cdots}$ and vice versa, which depends on the parity of the length.
We also note that the sequence $\overline{Z\cdots CD}$ in the right and the sequence $\overline{AB\cdots Z}$ in the left are sometimes absent.

In some cases, a $k$-support operator in \qh is generated only by three commutators.
This happens when the two leftmost or rightmost operators of the generated $k$-support operator are $XX$, $YY$, or $ZZ$.
An example is $ZXXXX$, which is generated only by the following three commutators:
\eq{
\ba{ccccc}{
Z&Z&&& \\
&Y&Y&& \\
&&Z&Z& \\
&&&X&X \\ \hline \hline
&&&Z& \\ \hline
},
\ \
\ba{ccccc}{
Z&Z&&& \\
&Y&Y&& \\
&&Z&Z& \\
&&&Y&Y \\ \hline \hline
&&&&Z \\ \hline
},
\ \
\ba{ccccc}{
&Y&Y&& \\
&&Z&Z& \\
&&&X&X \\
&&&Z& \\ \hline \hline
Z&Z&&& \\ \hline
}.
}

For brevity of explanation, below we shall treat only the case with odd $k$.
Extension to even $k$ is straightforward.
To prove that one of the remaining coefficients is zero, we consider the following sequence of operators for $k\geq 5$
\eqa{
\ba{ccccccc}{
\hz \y\z R^{k-3} &\threeskip& \x\hz\z R^{k-3} &\threeskip&\bz \y\z R^{k-4} \rplus \y &\threeskip& \\
\z\hz\y\z R^{k-4} &\threeskip& \z\x\hz\z R^{k-4}  &\threeskip&\z\bz\y\z R^{k-5} \rplus \x &\threeskip& \z \lplus \bz\y\z R^{k-4} \\
L^1 \z\hz\y\z R^{k-5} &\threeskip& L^1\z\x\hz\z R^{k-5}  &\threeskip&L^1\z\bz\y\z R^{k-6} \rplus \y &\threeskip& \x \lplus \z\bz\y\z R^{k-5} \\
L^2 \z\hz\y\z R^{k-6} &\threeskip& L^2\z\x\hz\z R^{k-6}  &\threeskip&L^2\z\bz\y\z R^{k-7} \rplus \x &\threeskip& \y \lplus L^1\z\bz\y\z R^{k-6} \\
\vdots &\threeskip& \vdots &\threeskip& \vdots &\threeskip& \vdots\\
L^n \z\hz\y\z R^{k-n-4} &\threeskip& L^n\z\x\hz\z R^{k-n-4}  &\threeskip&L^n\z\bz\y\z R^{k-n-5} \rplus \x &\threeskip& \y \lplus L^{n-1}\z\bz\y\z R^{k-n-4} \\
L^{n+1} \z\hz\y\z R^{k-n-5} &\threeskip& L^{n+1}\z\x\hz\z R^{k-n-5}  &\threeskip&L^{n+1}\z\bz\y\z R^{k-n-6} \rplus \y &\threeskip& \x \lplus L^{n}\z\bz\y\z R^{k-n-5} \\
\vdots &\threeskip& \vdots &\threeskip& \vdots &\threeskip& \vdots\\
L^{k-6} \z\hz\y\z R^2 &\threeskip& L^{k-6}\z\x\hz\z R^2  &\threeskip&L^{k-6}\z\bz\y\z R^1 \rplus \y &\threeskip& \x \lplus L^{k-7}\z\bz\y\z R^2 \\
L^{k-5} \z\hz\y\z R^1 &\threeskip& L^{k-5}\z\x\hz\z R^1  &\threeskip&L^{k-5}\z\bz\y\z  \rplus \x &\threeskip& \y \lplus L^{k-6}\z\bz\y\z R^1 \\
L^{k-4} \z\hz\y\z  &\threeskip& L^{k-4}\z\x\hz\z   &\threeskip&L^{k-4}\z\bz\y\rplus \z &\threeskip& \x \lplus L^{k-5}\z\bz\y\z \\
L^{k-3} \z\hz\y  &\threeskip& L^{k-3}\z\x\hz   &\threeskip&&\threeskip& \y \lplus L^{k-4}\z\bz\y
}
}{sequence-XYZh}
where operators in the same row generate the same $k$-support operator, and $n$ in the middle of rows is even.
Note that the case of $k=3$ has already been treated in the previous subsection.

We shall write down obtained relations on coefficients from rows with $n$ and $n+1$, which induce the following relation:
\balign{
h(-q_{L^n \z\y\z R^{k-n-4}}+q_{L^n\z\x\z R^{k-n-4} })-J^Xq_{L^n\z\bz\y\z R^{k-n-5}}+J^Y q_{L^{n-1}\z\bz\y\z R^{k-n-4}}&=0, \lb{genk-n-rel}  \\
h(-q_{L^{n+1} \z\y\z R^{k-n-5}}+q_{L^{n+1}\z\x\z R^{k-n-5} })+J^Yq_{L^{n+1}\z\bz\y\z R^{k-n-6}}-J^X q_{L^{n}\z\bz\y\z R^{k-n-5} }&=0. \lb{genk-n+1-rel}
}
Our tentative goal is to erase all of the coefficients of $k-1$-support operator and derive a relation on $\cX$, by multiplying a proper number and summing up the relations obtained by these rows.
Using the relation
\eqa{
q_{L^n \z\y\z R^{k-n-4}}=\frac{J^Y}{J^X}q_{L^n\z\x\z R^{k-n-4} },
}{XYZh-gen-qq-rel}
which follows from \eref{doubling-coefficient}, the resulting relation after the summation reads
\eq{
h\( \frac{J^X}{J^Y}-1\) k\cX^k \sigma({X,Z,R^{k-3}})J^{{\x\z R^{k-3}}}=0,
}
where $\cX^k $ is a constant introduced in \eref{doubling-coefficient}.
This relation implies the desired result.

\bigskip

Although the above proof can be confirmed by direct computation, it is worth clarifying the structure to determine the signs of coefficients in the above relations, which guarantees that we finally get a nontrivial relation on $\cX^k $.
In fact, an improper sequence induces a trivial relation $0=0$, and we can get no information on $\cX^k $, the constant introduced in \eref{doubling-coefficient} (see also the end of this subsection, where we present an example of an improper sequence).
We examine the minus sign of $q_{L^n\z\bz\y\z R^{k-n-5}}$ and the plus sign of $q_{L^{n-1}\z\bz\y\z R^{k-n-4}}$ in \eref{genk-n-rel} as examples.
The former sign $-1$ comes from that of the commutator $s(L^n\z\bz\y\z R^{k-n-5} \rplus \x)=-1$.
An important fact is that the former sign is equal to
\eqa{
s(L^n\z\bz\y\z R^{k-n-5} \rplus \x)=s(L^n\z\y\z R^{k-n-5} \rplus \x)=\frac{\sigma(L^n, Z, Y, Z, R^{k-n-4})}{\sigma(L^n, Z, Y, Z,  R^{k-n-5})},
}{s-remove-bz}
where the argument of the numerator is the generated operator by the commutator in the argument of the middle term: $L^n\z\bz\y\z R^{k-n-5} \rplus \x$.
The first equality of \eref{s-remove-bz} states that the following two commutators, which is the case with $n=0$ and $k=7$, have the same sign:
\eq{
\ba{cccccccc}{
&Z&Z&&&&& \\
&&Y&Y&&&& \\
&&&Z&Z&&& \\
&&&&X&X&& \\
&&&&&Y&Y& \\
&&Z&&&&& \\ \hline \hline
&&&&&&X&X \\ \hline
} \ \ \  {\rm and} \ \ \
\ba{cccccccc}{
&Z&Z&&&&& \\
&&Y&Y&&&& \\
&&&Z&Z&&& \\
&&&&X&X&& \\
&&&&&Y&Y& \\ \hline \hline
&&&&&&X&X \\ \hline
} ,
}
since the single $Z$ at this position does not affect the sign of the commutator.
Similarly, the sign of commutator $\y \lplus L^{n-1}\z\bz\y\z R^{k-n-4}$ is calculated as
\eq{
s(\y \lplus L^{n-1}\z\bz\y\z R^{k-n-4})=s(\y \lplus L^{n-1}\z\y\z R^{k-n-4}) =-\frac{\sigma(L^n,Z,Y,Z,R^{k-n-4})}{\sigma(L^{n-1},Z,Y,Z,R^{k-n-4})},
}
where the minus sign on the right-hand side comes from the minus sign in \eref{commute-minus}.

Keeping \eref{XYZh-gen-qq-rel} in mind, the relation on coefficients induced by the row with $n$ in \eref{sequence-XYZh} (which is equal to \eref{genk-n-rel}) is calculated as
\balign{
h\( \frac{J^Y}{J^X}-1\) \sigma(L^n,Z,Y,Z, R^{k-n-4}) J^{L^n \z\y\z R^{k-n-4}}\cdot \cX^k &+J^X\frac{\sigma(L^n,Z,Y,Z,R^{k-n-4})}{\sigma(L^n,Z,Y,Z, R^{k-n-5})} q_{L^n\z\bz\y\z R^{k-n-5}} \nt \\
&-J^Y \frac{\sigma(L^n,Z,Y,Z, R^{k-n-4})}{\sigma(L^{n-1},Z,Y,Z, R^{k-n-4})}q_{L^{n-1}\z\bz\y\z R^{k-n-4}}=0,
}
or equivalently
\balign{
h\( \frac{J^Y}{J^X}-1\) \cdot \cX^k +&\frac{J^X}{J^{L^n \z\y\z R^{k-n-4}}}\frac{q_{L^n\z\bz\y\z R^{k-n-5}}}{\sigma(L^n,Z,Y,Z, R^{k-n-5})} \nt \\
-&\frac{J^Y}{J^{L^n \z\y\z R^{k-n-4}}} \frac{q_{L^{n-1}\z\bz\y\z R^{k-n-4}}}{\sigma(L^{n-1},Z,Y,Z, R^{k-n-4})}=0.
}
In a similar manner to above, the relation on coefficients induced by the row with $n+1$ in \eref{sequence-XYZh} (which is equal to \eref{genk-n+1-rel}) is calculated as
\balign{
h\( \frac{J^Y}{J^X}-1\) \cdot \cX^k +&\frac{J^Y}{J^{L^{n+1} \z\y\z R^{k-n-5}}}\frac{q_{L^{n+1}\z\bz\y\z R^{k-n-6}}}{\sigma(L^{n+1},Z,Y,Z, R^{k-n-6})} \nt \\
-&\frac{J^X}{J^{L^{n+1} \z\y\z R^{k-n-5}}} \frac{q_{L^{n}\z\bz\y\z R^{k-n-5} }}{\sigma(L^{n},Z,Y,Z, R^{k-n-5})}=0.
}
Since $J^{L^n \z\y\z R^{k-n-4}}=J^{L^{n+1} \z\y\z R^{k-n-5}}$, by summing the above relations (without multiplying any number) all the coefficients of $k-1$-support operators cancel and the coefficient of $\cX^k $ is kept finite, which leads to a relation in the form of $\cX^k \times (\text{finite number}) =0$.

\bigskip

For a better understanding, we here present an example of an improper sequence that conveys a trivial relation $0=0$.
An example with $k=4$ is
\eq{
\ba{ccccccc}{
\hz \y\z\x &\threeskip& \x\hz\z\x &\threeskip&\bz \y\z  \rplus \x &\threeskip& \\
\hz\x\y\z &\threeskip& \y\x\hz\z   &\threeskip&\y\bz\y \rplus \z &\threeskip& \y \lplus \bz\y\z \\
\hz\x\y\x &\threeskip& \y\x\y\hz   &\threeskip&\y\bz\y \rplus \x &\threeskip& \y \lplus \bz\y\x \\
\hz \y\x\z &\threeskip& \x\y\hz\z &\threeskip&\bz \y\x  \rplus \z &\threeskip&
}
}
which produces four relations
\balign{
hq_{\y\z\x}+hq_{\x\z\x}+J^Xq_{\bz \y\z}&=0, \\
-hq_{\x\y\z}+hq_{\y\x\z}+J^Zq_{\y\bz\y}+J^Yq_{\bz\y\z}&=0, \\
-hq_{\x\y\z}+hq_{\y\x\y}-J^Xq_{\y\bz\y}+J^Yq_{\bz\y\x}&=0, \\
hq_{\y\x\z}-hq_{\x\y\z}-J^Zq_{\bz \y\x}&=0.
}
Erasing $q_{\bz \y\z}$, $q_{\y\bz\y}$, and $q_{\bz \y\x}$, we find a trivial relation $0=0$ and cannot extract any information on $\cX^4$.


\section{Heisenberg model with next-nearest-neighbor interaction}\lb{s:NNN}

\subsection{Commutators generating $k+2$-support operators}\lb{s:NNN-step1}

The proof idea is again that presented in \sref{proof-idea}.
Namely, by supposing the conservation of a $k$-support operator $Q$, we examine the commutation relation in detail and show that $q_{\bsA ^k}=0$ for all $\bsA^k$.

In the case of the NNN-XYZ model, a commutator of a $k$-support operator and the Hamiltonian can generate at most $k+2$-support operators.
Therefore, we first consider the case that the commutator generates $k+2$-support operators.

A $k+2$-support operator in \qh is generated only by a commutator such that the next-nearest-neighbor interaction term ($XIX$, $YIY$, and $ZIZ$) acts on the left end or right end of a $k$-support operator.
The following two types of commutators serve as examples:
\eq{
\ba{cccccc}{
A&B&\cdots&C&& \\
&&&X&I &X \\ \hline
A&B&\cdots&D&I &X
}
, \ \
\ba{cccccc}{
&&E&\cdots&F&G \\
Y&I &Y&&& \\ \hline
Y&I &H&\cdots&F&G
}.
}

To explain our result, we introduce several symbols in addition to \sref{symbol1}.
We first introduce {\it extended-doubling-products} by using a tilde symbol as $\tx=X_iX_{i+2}=XIX$, $\ty=YIY$, and $\tz=ZIZ$.
Similarly to the doubling-product operator, we introduce an {\it extended-doubling-product operator} which is an operator expressed as, e.g., $\tx\ty\tx\tz$ with divesting its phase factor.
Here, we promise that a neighboring extended-doubling-product has its support with a two-site shift.
The aforementioned extended-doubling-product operator $\tx\ty\tx\tz$, for example, reads
\eq{
\tx\ty\tx\tz=|(X_iX_{i+2})\cdot (Y_{i+2}Y_{i+4})\cdot(X_{i+4}X_{i+6})\cdot(Z_{i+6}Z_{i+8})|=\ba{ccccccccc}{
X&I&X&&&&&& \\
&&Y&I&Y&&&& \\
&&&&X&I&X&& \\
&&&&&&Z&I&Z \\ \hline \hline
}
=XIZIZIYIZ.
}
Here, the double horizontal line represents the column expression introduced in \sref{symbol1}, where we take products in the vertical direction under the rule of the signless products.

Now we employ a similar argument to \sref{XYZh-step1}, leading to a constraint similar to the doubling-product operator in the XYZ+h model, where the extended-doubling-product operator plays the role of the doubling-product operator in \sref{XYZh-step1}.
Precisely, only $k$-support extended-doubling-product operators may have a nonzero coefficient in $Q$, and other $k$-support operators have zero coefficients.
We shall explain the latter point briefly.
Consider operator $XIZIZZYIXIY$ in the case of $k=11$.
Then, this operator can be expressed as
\eq{
XIZIZZYIXIY=
\ba{ccccccccccc}{
X&I&X&&&&&&&& \\
&&Y&I&Y&&&&&& \\
&&&&X&Z&X&&&& \\
&&&&&&Z&I&Z&& \\
&&&&&&&&Y&I&Y \\ \hline \hline
},
}
where a ``defect" $XZX$ is inserted.
This defect lies in the following series of pairs
\eq{
\ba{ccccccccccc}{
X&I&X&&&&&&&& \\
&&Y&I&Y&&&&&& \\
&&&&X&Z&X&&&& \\
&&&&&&Z&I&Z&& \\
&&&&&&&&Y&I&Y \\ \hline \hline
}
\lr
\ba{ccccccccccc}{
Y&I&Y&&&&&&&& \\
&&X&Z&X&&&&&& \\
&&&&Z&I&Z&&&& \\
&&&&&&Y&I&Y&& \\
&&&&&&&&Z&I&Z \\ \hline \hline
}
\lr
\ba{ccccccccccc}{
X&Z&X&&&&&&&& \\
&&Z&I&Z&&&&&& \\
&&&&Y&I&Y&&&& \\
&&&&&&Z&I&Z&& \\
&&&&&&&&Y&I&Y \\ \hline \hline
}.
}
However, the last operator $XZYIXIXIXIY$ cannot form a pair because the two leftmost operators are $XZ\cdots$, not in the form of $XI\cdots$:
\eq{
\ba{ccccccccccccc}{
X&Z&Y&I&X&I&X&I&X&I&Y&& \\
&&&&&&&&&&Z&I&Z \\ \hline
X&Z&Y&I&X&I&X&I&X&I&X&I&X
},
\hspace{15pt}
\ba{ccccccccccccc}{
&&?&?&?&?&?&?&?&?&?&?&? \\
X&Z&?&&&&&&&&&& \\ \hline
X&Z&Y&I&X&I&X&I&X&I&X&I&X
}.
}
This fact implies
\eq{
q_{XZYIXIXIXIY}=0,
}
and hence the initial operator $XIZIZZYIXIY$ also has zero coefficient:
\eq{
q_{XIZIZZYIXIY}=0.
}
In general, if two leftmost operators are not one of $XI\cdots$, $YI\cdots$, or $ZI\cdots$, or two rightmost operators are not one of $\cdots IX$, $\cdots IY$, or $\cdots IZ$, then this operator cannot form a pair.
Thus, arguments similar to \sref{XYZh-step1} (proof of Lemma 1) confirm that if a $k$-support operator is not an extended-doubling-product operator, then by removing extended-doubling-products from left and adding extended-doubling-products to right repeatedly, we arrive at an operator which cannot form a pair, resulting a zero coefficient.

\blm{\lb{l:NNN-k}
Let $Q$ be a $k$-support conserved quantity expanded as \eqref{Qform}.
Then, its coefficients of a $k$-support operator $\bsA\in \calP^k$ should be expressed as
\eqa{
q_{\bsA}=\cN^k\cdot \sigma(B^1,B^2,\ldots , B^{(k-1)/2}) \prod_{i=1}^{(k-1)/2} J_2^{B^i}
}{NNN-doubling-coefficient-1}
with a common constant $\cN^k$ if $\bsA$ is an extended-doubling-product operator written as $\bsA=\prod_{i=1}^{(k-1)/2} \widetilde{B}^i$, and zero otherwise.
}

Clearly, a $k$-support conserved quantity with even $k$ vanishes.

\subsection{Commutators generating $k+1$-support operators}\lb{s:NNN-step2}

We next consider the case that a commutator generates $k+1$-support operators.

We notice that only the following two commutators generate a $k+1$-support operator $XIZI\cdots IYX$:
\eq{
\ba{ccccccccc}{
X&I&X&&&&&& \\
&&Y&I&Y&&&& \\
&&&&\ddots&&&& \\
&&&&&Z&I&Z& \\ \hline \hline
&&&&&&&X&X \\ \hline
}
, \ \ \ \
\ba{ccccccccc}{
&&Y&I&Y&&&& \\
&&&&\ddots&&&& \\
&&&&&Z&I&Z& \\
&&&&&&&X&X \\ \hline \hline
X&I&X&&&&&& \\ \hline
}.
}
Using symbols introduced in \sref{symbol1} and \sref{symbol2}, the above two commutators are expressed as
\eq{
\tx\ty \cdots \tz \rplus\x , \ \ \ \tx \lplus\ty\cdots \tz\x ,
}
which implies the following relation of coefficients:
\eq{
J_1^Xq_{\tx\ty \cdots \tz}-J_2^Xq_{\ty\cdots \tz\x}=0.
}
This relation connects the coefficient of a $k$-support operator and that of a $k-1$-support operator.

We can connect two coefficients of $k-1$-support operators, e.g., $q_{\ty\tx\cdots \tz\x}$ and $q_{\tx\cdots \tz\x\ty}$, by considering the following two commutators:
\eq{
\ba{ccccccccccc}{
Y&I&Y&&&&&&&& \\
&&X&I&X&&&&&& \\
&&&&\ddots&&&&&& \\
&&&&&Z&I&Z&&& \\
&&&&&&&X&X&& \\ \hline \hline
&&&&&&&&Y&I&Y \\ \hline
}
,\ \ \
\ba{ccccccccccc}{
&&X&I&X&&&&&& \\
&&&&\ddots&&&&&& \\
&&&&&Z&I&Z&&& \\
&&&&&&&X&X&& \\
&&&&&&&&Y&I&Y \\ \hline  \hline
Y&I&Y&&&&&&&& \\ \hline
},
}
which implies
\eq{
J^Y_2q_{\ty\tx\cdots \tz\x}+J^Y_2q_{\tx\cdots \tz\x\ty}=0.
}
This observation suggests that the coefficient of a $k-1$-support operator written as the product of $(k-3)/2$ extended-doubling-products and one doubling-product, e.g., $\tx\ty\tx\z\tx\tz$, is connected to the coefficients of $k$-support operators.
For example, $\tx\ty\tx\z\tx\tz$ forms a pair as
\eq{
\tx\ty\tx\z\tx\tz \lr \ty\tx\ty\tx\z\tx \lr \tx\ty\tx\ty\tx\z \lr \ty\tx\ty\tx\ty\tx ,
}
where the first three operators are $k-1$-support and the last one is $k$-support.

\bigskip

An important fact is that a $k-1$-support operator except for the above form has a zero coefficient.
We first demonstrate that a $k+1$-support operator is generated by at most two commutators.
At first glance, a $k+1$-support operator can be generated by four commutators:
\balign{
&\tb\lplus (k-1\text{-support operator}), \\
&\b \lplus (k\text{-support operator}), \\
& (k-1\text{-support operator})\rplus \tb,\\
& (k\text{-support operator})\rplus \b,
}
whose column expressions are
\eqa{
\ba{cccccccc}{
&&?&?&?&?&?&? \\
B&I&B&&&&& \\ \hline
}, \hspace{10pt}
\ba{cccccccc}{
&?&?&?&?&?&?&? \\
B&B&&&&&& \\ \hline
}, \hspace{10pt}
\ba{cccccccc}{
?&?&?&?&?&?&& \\
&&&&&B&I&B \\ \hline
}, \hspace{10pt}
\ba{cccccccc}{
?&?&?&?&?&?&?& \\
&&&&&&B&B \\ \hline
}.
}{NNN-k-1-4}
Here, we excluded the possibility of commutators between $\tb$ and $k$-support operator such that $\tb$ nontrivially acts on the second left (or right) site of the $k$-support operator, since these column expressions are
\eq{
\ba{cccccccc}{
&?&?&?&?&?&?&? \\
B&I&B&&&&& \\ \hline
}, \hspace{10pt}
\ba{cccccccc}{
?&?&?&?&?&?&?& \\
&&&&&B&I&B \\ \hline
},
}
but Lemma.~\ref{l:NNN-k} tells that all $k$-support operator with nonzero coefficients has an identity operator at the second left site and the second right site, implying that these commutators vanish.

From \eref{NNN-k-1-4}, we see that only two of these four commutators generate a given $k+1$-operator.
To see this, we focus on the operators at the second left and second right sites.
If the operator at the second left site is an identity operator $I$, the second commutator in \eref{NNN-k-1-4} never generates this operator, and if  the operator at the second left site is a Pauli operator $X$, $Y$ or $Z$, the first commutator in \eref{NNN-k-1-4} never generates this operator.
A similar argument holds for the second right site.
In summary, only two commutators in \eref{NNN-k-1-4} generate a single $k+1$-support operator.

We now demonstrate how a $k-1$-support operator which is not a product of many extended-doubling-products and single doubling-product is shown to have zero coefficient.
Consider the case of $k=11$ and $XIZIZYYIYZ$ as an example.
The operator $XIZIZYYIYZ$ can be expressed in the column expression as
\eq{
XIZIZYYIYZ=
\ba{cccccccccc}{
X&I&X&&&&&&& \\
&&Y&I&Y&&&&& \\
&&&&X&X&&&& \\
&&&&&Z&Z&&& \\
&&&&&&X&I&X& \\
&&&&&&&&Z&Z \\ \hline \hline
}.
}
Note that there are one $XX$ and two $ZZ$'s.
These operators appear in the following series of pairs
\eq{
\ba{cccccccccc}{
X&I&X&&&&&&& \\
&&Y&I&Y&&&&& \\
&&&&X&X&&&& \\
&&&&&Z&Z&&& \\
&&&&&&X&I&X& \\
&&&&&&&&Z&Z \\ \hline \hline
}\lr
\ba{cccccccccc}{
Y&I&Y&&&&&&& \\
&&X&X&&&&&& \\
&&&Z&Z&&&&& \\
&&&&X&I&X&&& \\
&&&&&&Z&Z&& \\
&&&&&&&X&I&X \\ \hline \hline
}\lr
\ba{cccccccccc}{
X&X&&&&&&&& \\
&Z&Z&&&&&&& \\
&&X&I&X&&&&& \\
&&&&Z&Z&&&& \\
&&&&&X&I&X&& \\
&&&&&&&Y&I&Y \\
\hline \hline
}\lr
\ba{ccccccccccc}{
Z&Z&&&&&&&&& \\
&X&I&X&&&&&&& \\
&&&Z&Z&&&&&& \\
&&&&X&I&X&&&& \\
&&&&&&Y&I&Y&& \\
&&&&&&&&X&I&X \\
\hline \hline
},
}
where the first three are $k-1$-support operators while the last one is a $k$-support operator.
The last operator, $ZYIYYIZIZIX$, is not an extended-doubling-product operator and thus has zero coefficient, $q_{ZYIYYIZIZIX}$, which implies that the first operator also has zero coefficient, $q_{XIZIZYYIYZ}=0$.

Following similar arguments to above, we find the following result:

\blm{
Let $Q$ be a $k$-support conserved quantity expanded as \eqref{Qform}.
Then, its coefficients of a $k-1$-support operator $\bsA\in \calP^{k-1}$ should be expressed as
\eqa{
q_{\bsA}=\cN^k\cdot \sigma(B^1, B^2,\ldots , B^{(k-1)/2}) J^{B^m}_1\prod_{i=1, i\neq m}^{(k-1)/2} J_2^{B^i}
}{NNN-doubling-coefficient-2}
with a common constant $\cN^k$ to \eref{NNN-doubling-coefficient-1} if $\bsA$ is an operator written as $\bsA=\tb_1\tb_2\cdots \tb_{m-1} \b_m \tb_{m+1}\cdots \tb_{(k-1)/2}$, and $q_{\bsA}=0$ otherwise.
}

\subsection{Commutators generating $k$-support operators}\lb{s:NNN-step3}

We finally consider the case that a commutator generates a $k$-support operator.
Similarly to \sref{XYZh-step2}, we introduce symbols $\hlz$ and $\hrz$, representing commutation relations with $ZZ$ at this position.
We express, for example, these commutators
\eq{
\ba{ccccc}{
Y&I&Y&& \\
&&Z&I&Z \\ \hline \hline
&&Z&Z& \\ \hline
}
, \ \ \
\ba{ccccc}{
Y&I&Y&& \\
&&Z&I&Z \\ \hline \hline
&Z&Z&& \\ \hline
}
}
as
\eq{
\ty \hrz \tz , \ \ \ \ty \hlz \tz
}
respectively
We also introduce $\blz$ and $\brz$, representing multiplication of $ZZ$ at this position.
We express, for example, these commutators
\eq{
\ba{ccccc}{
Y&I&Y&& \\
&&Z&Z& \\ \hline \hline
&&Z&I&Z \\ \hline

}
, \ \ \
\ba{ccccc}{
Y&I&Y&& \\
&Z&Z&& \\ \hline \hline
&&Z&I&Z \\ \hline
}
}
as
\eq{
\ty \brz \rplus\tz , \ \ \ \ty \blz \rplus\tz
}
respectively.
Remark that the relative position of the vertical bar or the vertical arrow and symbol $ZZ$ represents which $Z$ in the doubling-product $ZZ$ acts nontrivially.
The symbol $\blz$ (resp. $\brz$) represents that the right (resp. left) $Z$ in $ZZ$ acts.
We also note that although both $\ty \brz$ and $\ty \z$ represent the same operator $YIXZ$ (i.e., $\ty \brz =\ty \z=YIXZ$), we promise the following rule: $\ty \z \rplus \ty$ means that $\ty$ acts on the right end of $\z$ and generates 6-local operator $YIXXIY$, while $\ty \brz \rplus\tz$ means that $\tz$ acts on the right end of $\ty$ and generates 5-local operator $YIZZY$.
Two commutators $\ty \brz\rplus \ty$ and $\ty \z\rplus \ty$ are represented as
\eq{
\ty \brz\rplus \ty=
\ba{cccccc}{
Y&I&Y&&& \\
&&Z&Z&& \\ \hline \hline
&&&Y&Y \\ \hline
},
\hspace{25pt}
\ty \z\rplus \ty=
\ba{ccccc}{
Y&I&Y&& \\
&&Z&Z& \\ \hline \hline
&&Y&Y \\ \hline
}.
}

We further introduce symbols which represent alternating $\tx$ and $\ty$ defined as
\balign{
\tL^{2n}&:=\underbrace{\ty\tx\cdots \ty\tx}_{n \text{ copies of }\ty\tx}, \\
\tL^{2n+1}&:=\tx\underbrace{\ty\tx\cdots \ty\tx}_{n \text{ copies of }\ty\tx}, \\
\tR^{2n}&:=\underbrace{\tx\ty\cdots \tx\ty}_{n \text{ copies of }\tx\ty}, \\
\tR^{2n+1}&:=\underbrace{\tx\ty\cdots \tx\ty}_{n \text{ copies of }\tx\ty}\tx.
}
Similarly to the previous section (\sref{symbol2}), when we consider the sign $\sigma$, we promise that $\tL$ and $\tR$ mean $Y,X,Y,X\ldots$ and $X,Y,X,\ldots$, respectively.

\bigskip

Now we construct a sequence of commutators, with which we can demonstrate $\cN^k=0$.
For the brevity of explanation, we only treat the case of $k\equiv 3\mod 4$ .
The extension to the case of $k\equiv 1\mod 4$ is straightforward.
We express $k=4r+3$ and consider the following sequence:
\eqa{
\ba{ccccccc}{
\tL^{2r} \hrz \tz &\threeskip& \tL^{2r} \brz \rplus\tz &&&\threeskip&  \ty \lplus \tL^{2r-1} \brz \tz \\
\tL^{2r-1} \hrz \tz\tR^1 &\threeskip&&& \tL^{2r-1} \brz \tz \rplus\tx &\threeskip& \tx \lplus \tL^{2r-2} \brz \tz\tR^1 \\
\tL^{2r-2} \hrz \tz\tR^2 &\threeskip&&& \tL^{2r-2} \brz \tz\tR^1 \rplus\ty &\threeskip& \ty \lplus \tL^{2r-3} \brz \tz\tR^2 \\
\vdots &\threeskip&&& \vdots &\threeskip& \vdots \\
\tL^{2r-n} \hrz \tz\tR^n &\threeskip&&& \tL^{2r-n} \brz \tz \tR^{n-1} \rplus\ty &\threeskip& \ty \lplus \tL^{2r-n-1} \brz \tz\tR^n \\
\tL^{2r-n-1} \hrz \tz\tR^{n+1} &\threeskip&&& \tL^{2r-n-1} \brz \tz\tR^n \rplus\tx &\threeskip& \tx \lplus \tL^{2r-n-2} \brz \tz\tR^{n+1} \\
\vdots &\threeskip&&& \vdots &\threeskip& \vdots \\
\tL^2 \hrz \tz\tR^{2r-2} &\threeskip&&& \tL^2 \brz \tz\tR^{2r-3} \rplus\ty &\threeskip& \ty \lplus \tL^1 \brz \tz\tR^{2r-2} \\
\tL^1 \hrz \tz\tR^{2r-1} &\threeskip&&& \tL^1 \brz \tz\tR^{2r-2}\rplus \tx &\threeskip& \\
}
}{sequence-NNN}
where $n$ is even.
The leftmost column has commutators between a $k$-body operator and 2-body operator $ZZ$ (in the Hamiltonian), the second left column has a commutator (in the first row) between a $k-1$-body operator and a 3-body operator $\tz$ in the Hamiltonian, and the two right columns show commutators between a $k-2$-body operator and a 3-body operator ($\tx$ or $\ty$) in the Hamiltonian.

We put two remarks:
First, each operator corresponding to each row is generated only by a single commutator between a $k$-body operator and 2-body operator $\hrz$, because generating the operators by $\hlz$, the corresponding $k$-support operator is not an extended-doubling-product operator.
Second, the last law generating $YI YZY\cdots$ has only two elements, because we cannot obtain this operator by a commutator in the form of $\ty \lplus (k-2 \text{-support operator})$.

\bigskip

Now we examine their signs.
First, all the commutators in the leftmost column has the plus sign, which follows from
\eq{
s(\tL^m \hrz \tz \tR^{m'})=s([Y,Z])=+1.
}
Next, to compute the signs of commutators in the second right column we notice
\eqa{
s( \tL^{2r-n} \brz \tz \tR^{n-1} \rplus\ty)=s( \tL^{2r-n}\tz \tR^{n-1} \rplus\ty),
}{NNN-right-sign}
which holds for the same reason as the first equality of \eref{s-remove-bz}.
An example of this fact can be seen by using the column expression as
\eq{
s\( \ba{ccccccccccc}{
Y&I&Y&&&&&&&& \\
&&X&I&X&&&&&& \\
&&&&Z&I&Z&&&& \\
&&&&&&X&I&X&& \\
&&&&Z&Z&&&&& \\ \hline \hline
&&&&&&&&Y&I&Y \\ \hline
}\)
=
s\( \ba{ccccccccccc}{
Y&I&Y&&&&&&&& \\
&&X&I&X&&&&&& \\
&&&&Z&I&Z&&&& \\
&&&&&&X&I&X&& \\ \hline \hline
&&&&&&&&Y&I&Y \\ \hline
}\)
}
Using the same technique as \eref{commute-plus}, the right-hand side of \eref{NNN-right-sign} is computed as
\eq{
s( \tL^{2r-n}\tz \tR^{n-1} \rplus\ty)=\frac{\sigma(\tL^{2r-n},Z,\tR^{n})}{\sigma(\tL^{2r-n},Z, \tR^{n-1})}.
}
Similarly, we compute the signs of commutators in the rightmost column as
\eq{
s(\ty \lplus \tL^{2r-n-1} \brz \tz\tR^n)=s(\ty \lplus \tL^{2r-n-1} \tz\tR^n)=-\frac{\sigma(\tL^{2r-n},Z,\tR^n)}{\sigma(\tL^{2r-n-1},Z,\tR^n)},
}
where the minus sign comes from the same reason as the minus sign in \eref{commute-minus}:
\eq{
s\( \ba{ccccccccccc}{
&&X&I&X&&&&&& \\
&&&&Z&I&Z&&&& \\
&&&&&&X&I&X&& \\
&&&&&&&&Y&I&Y \\ \hline \hline
Y&I&Y&&&&&&&& \\ \hline
}
\) = - \frac{\sigma(Y,X,Z,X,Y)}{\sigma(X,Z,X,Y)}.
}

Hence, the relation obtained from the $n+1$-th row (except $n=0$) reads
\balign{
J^Z_1 \sigma(\tL^{2r-n},Z,\tR^n)J^{\tL^{2r-n}\tz\tR^n}\cdot \cN^k+&J^Y_2\frac{\sigma(\tL^{2r-n},Z,\tR^{n})}{\sigma(\tL^{2r-n},Z, \tR^{n-1})}q_{ \tL^{2r-n}\brz \tz \tR^{n-1}} \nt \\
-&J^Y_2\frac{\sigma(\tL^{2r-n},Z,\tR^n)}{\sigma(\tL^{2r-n-1},Z,\tR^n)}q_{\tL^{2r-n-1} \brz \tz\tR^n}=0, \lb{NNN-step3-mid1}
}
which is equivalent to
\eqa{
J^Z_1\cdot \cN^k+\frac{J^Y_2}{J^{\tL^{2r-n}\tz\tR^n}}\frac{q_{ \tL^{2r-n}\brz \tz \tR^{n-1}}}{\sigma(\tL^{2r-n},Z, \tR^{n-1})}-\frac{J^Y_2}{J^{\tL^{2r-n}\tz\tR^n}}\frac{q_{\tL^{2r-n-1} \brz \tz\tR^n}}{\sigma(\tL^{2r-n-1},Z,\tR^n)}=0.
}{NNN-step3-mid1b}
In a similar manner to above, the relation on coefficients obtained from the $n+2$-th row (except $n=2r-2$) reads
\eqa{
J^Z_1\cdot \cN^k+\frac{J^X_2}{J^{\tL^{2r-n-1}\tz\tR^{n+1}}}\frac{q_{ \tL^{2r-n-1}\brz \tz \tR^{n}}}{\sigma(\tL^{2r-n-1},Z, \tR^{n})}-\frac{J^X_2}{J^{\tL^{2r-n-1}\tz\tR^{n+1}}}\frac{q_{\tL^{2r-n-2} \brz \tz\tR^{n+1}}}{\sigma(\tL^{2r-n-2},Z,\tR^{n+1})}=0.
}{NNN-step3-mid2}
In addition, the last row implies
\eq{
J^Z_1\cdot \cN^k+\frac{J^X_2}{J^{\tL^{2r-1}\tz\tR}}\frac{q_{ \tL^{2r-1}\brz \tz}}{\sigma(\tL^{2r-1},Z)}=0.
}
Noticing that for even $n'$ and odd $n''$
\eq{
\frac{J^X_2}{J^Y_2}J^{\tL^{2r-n'}\tz\tR^{n'}}=J^{\tL^{2r-n''}\tz\tR^{n''}}
}
is satisfied, we find that by summing these relations (Eqs.~\eqref{NNN-step3-mid1b} and \eqref{NNN-step3-mid2}) from $n=0$ to $n=2r-1$ all of coefficients of $k-2$-support operators cancel and only the coefficients on $k$-support and $k-1$-support operators remain.

We shall finally derive an explicit relation on $\cN^k$.
To this end, we clarify the relation on the $k-1$-support operator (i.e., the top row).
Noticing that the sign of $\tL^{2r}\brz$ and $\tL^{2r} \brz\tz$ are given by $\sigma(\tL^{2r}, Z)$ and $\sigma(\tL^{2r})$, we compute this relation as
\eqa{
J^Z_1 \sigma(\tL^{2r},Z)J^{\tL^{2r}\tz}\cdot \cN^k+J^Z_2 s(\tL^{2r} \brz \rplus\tz) \cN^k\cdot \sigma(\tL^{2r},Z)J^{\tL^{2r}\brz } -J^Y_2\frac{\sigma(\tL^{2r},Z)}{\sigma(\tL^{2r-1},Z)}q_{\tL^{2r-1} \brz \tz}=0,
}{NNN-step3-n0}
where $\cN^k$ is a constant introduced in \eref{NNN-doubling-coefficient-1}.
The quantity $s(\tL^{2r} \brz \rplus\tz)\sigma(\tL^{2r},Z)$ in the second term is computed as
\eq{
s(\tL^{2r} \brz \rplus\tz)\sigma(\tL^{2r},Z)=s([Y,Z])\sigma(\tL^{2r},Z)=\sigma(\tL^{2r},Z).
}
Inserting this, the sum of the relations (Eqs.~\eqref{NNN-step3-mid1b} and \eqref{NNN-step3-mid2})  from $n=0$ to $n=2r-1$ leads to
\eq{
2r J^Z_1 \cN^k=0,
}
which implies that all the coefficients of $k$-support operators in $Q$ is zero.
This completes the proof.

\subsection{Extension to systems without translation invariance}\lb{s:extend-NNN}

By a careful examination of the above derivation, we find that all the $ZZ$ term, both commutators ($\hrz$ and $\hlz$) and multiplications ($\brz$ and $\blz$), in the sequence \eqref{sequence-NNN} acts on the same two sites.
This means that even if the nearest-neighbor interaction terms are position-dependent, i.e., the Hamiltonian is given by
\eqa{
H=\sum_{i=1}^L  [ J_{1,i}^X X_i X_{i+1}+J_{1,i}^Y Y_iY_{i+1}+J_{1,i}^ZZ_iZ_{i+1}] + \sum_{i=1}^L[ J_2^X X_i X_{i+1}+J_2^Y Y_iY_{i+1}+J_2^ZZ_iZ_{i+1}]
}{H-nonshift}
with position-dependent coefficients $J_{1,i}^X$, $J_{1,i}^Y$, and $J_{1,i}^Z$, our proof still works, and the absence of local conserved quantity can be shown as long as one of $J_{1,i}^X$, $J_{1,i}^Y$, and $J_{1,i}^Z$ is nonzero at some $i$.

Since Lemma 2 is shown by using only the next-nearest-neighbor interaction terms, the same statement as Lemma 2 holds for the Hamiltonian \eqref{H-nonshift}.
On the other hand, by expanding $Q$ by $Q=\sum_{l=1}^{k} \sum_{\bsA^l\in \calP^l} \sum_{i=1}^L q_{\bsA^l,i}\bsA^l_i$, \eref{NNN-doubling-coefficient-2} in Lemma 3 should be replaced by
\eq{
q_{\bsA^{k-1},j}=\cN^k\cdot \sigma(B^1, B^2,\ldots , B^{(k-1)/2}) J^{B^m}_{1, j+2(m-1)} \prod_{i=1, i\neq m}^{(k-1)/2} J_2^{B^i}.
}
In our final step, the key relation \eqref{NNN-step3-mid1} is replaced by
\balign{
J^Z_{1,i^*} \sigma(\tL^{2r-n},Z,\tR^n)J^{\tL^{2r-n}\tz\tR^n}\cdot \cN^k&+J^Y_2\frac{\sigma(\tL^{2r-n},Z, \tR^{n})}{\sigma(\tL^{2r-n},Z, \tR^{n-1})}q_{ \tL^{2r-n}\brz \tz \tR^{n-1},i^*} \nt \\
&-J^Y_2\frac{\sigma(\tL^{2r-n},Z,\tR^n)}{\sigma(\tL^{2r-n-1},Z,\tR^n)}q_{\tL^{2r-n-1} \brz \tz\tR^n,i^*}=0,
}
where $i^*$ is a fixed site independent of $n$.
The coefficients $q_{ \tL^{2r-n}\brz \tz \tR^{n-1},i^*}$ depend on the position, which is determined by the second subscript $i^*$ referring to the position of $ZZ$.
Following the same argument, we obtain counterpart relations to Eqs.~\eqref{NNN-step3-mid1b} and \eqref{NNN-step3-mid2} as
\balign{
J^Z_{1,i^*}\cdot \cN^k+\frac{J^Y_2}{J^{\tL^{2r-n}\tz\tR^n}}\frac{q_{ \tL^{2r-n}\brz \tz \tR^{n-1},i^*}}{\sigma(\tL^{2r-n},Z, \tR^{n-1})}-\frac{J^Y_2}{J^{\tL^{2r-n},Z,\tR^n}}\frac{q_{\tL^{2r-n-1} \brz \tz\tR^n,i^*}}{\sigma(\tL^{2r-n-1},Z,\tR^n)}&=0, \\
J^Z_{1,i^*}\cdot \cN^k+\frac{J^X_2}{J^{\tL^{2r-n-1}\tz\tR^{n+1}}}\frac{q_{ \tL^{2r-n-1}\brz \tz \tR^{n},i^*}}{\sigma(\tL^{2r-n-1},Z, \tR^{n})}-\frac{J^X_2}{J^{\tL^{2r-n-1}\tz\tR^{n+1}}}\frac{q_{\tL^{2r-n-2} \brz \tz\tR^{n+1},i^*}}{\sigma(\tL^{2r-n-2},Z,\tR^{n+1})}&=0,
}
where the coefficient of the first term $J^Z_1$ is replaced by $J^Z_{1,i^*}$, and others are the same.
Moreover, the counterpart of \eref{NNN-step3-n0} reads
\eq{
J^Z_{1,i^*} \sigma(\tL^{2r},Z)J^{\tL^{2r}\tz}\cdot \cN^k+J^Z_2 s(\tL^{2r} \brz \rplus\tz) \cN^k\cdot \sigma(\tL^{2r},Z)J^{\tL^{2r}\brz,i^* } -J^Y_2\frac{\sigma(\tL^{2r},Z)}{\sigma(\tL^{2r-1},Z)}q_{\tL^{2r-1} \brz \tz,i^*}=0.
}
Combining them and following the same argument as above, we obtain
\eq{
2r J^Z_{1,i^*}\cN^k=0.
}
Since $i^*$ is arbitrary, this means that $\cN^k=0$ if $J^Z_i\neq 0$ at some $i$.

Owing to this extension, a quantum spin chain model proposed by Shastry and Sutherland~\cite{SS81} is covered\fn{
We note that this model is different from the famous two-dimensional model, which is frequently called the Shastry-Sutherland model.
}.
The Shastry-Sutherland model is the XYZ chain with next-nearest-neighbor interaction, whose next-nearest-neighbor interaction is shift-invariant while its nearest-neighbor interaction is invariant by two-site shift.
Our argument demonstrates that the Shastry-Sutherland model has no nontrivial local conserved quantity.

\section{Discussion}

We have rigorously shown that the anisotropic Heisenberg chain (XYZ chain) with next-nearest-neighbor interaction has no local conserved quantity.
The Hamiltonian we treat includes important models as a special case, the Majumdar-Ghosh model, the Shastry-Sutherland model, and other zigzag spin chains, which are prominent examples of frustration-free systems.
Our computation on the signs of commutators and coefficients is systematic, and with this detailed analysis, we clarify the reason why the proposed sequences in Eqs.~\eqref{sequence-XYZh} and \eqref{sequence-NNN} work to exclude the possibility of a nontrivial local conserved quantity.

These key sequences of commutators follow the structure that additional terms (Z magnetic field in the XYZ+h-model and $ZZ$ interaction term in the NNN-XYZ model) settle at the right end at first, and they move to the left end at last.
As far as we seek, this structure is necessary to obtain a nontrivial relation for the coefficient $c$ in the expansion \eqref{Qform} ($\cX$ and $\cN$).
Otherwise, a sequence has the additional term at the right end at both the first and last of the sequence, where the coefficient of $c$ vanishes and we obtain a trivial relation $0=0$.
With noticing that the presented proof technique is also useful to determining local conserved quantities in some integrable systems~\cite{NF20, YF23, Fuk23, Fuk24}, we expect that the distinction of integrability and non-integrability reduces to whether a {\it good} sequence exists or not, which will shed new light on the integrability.

As presented, our proof method is valid for systems not only with nearest-neighbor interaction but also with next-nearest-neighbor interaction, which suggests that our method applies to systems with longer interactions.
In particular, if the longest interaction is the Heisenberg type, then following a similar argument to \sref{XYZh-step1} and \sref{NNN-step1}, a possible form of $k$-support operators with nonzero coefficient is limited to extended-doubling-product-type operators by replacing $\tx=XIX$, $\ty=YIY$, and $\tz=ZIZ$ by $XII\cdots IX$, $YII\cdots IY$, and $ZII\cdots IZ$, respectively.
Although further analyses depend on the model in consideration, we strongly expect that many complex models can be proven to be indeed non-integrable by this method.

\newpage

\appendix

\begin{center}
{\bf \Large Appendix}
\end{center}

\section{Translation invariance of local conserved quantity}
In the main part, we restrict a possible form of a local conserved quantity to the shift-invariant form \eqref{Qform}.
In this Appendix, we show that this does not lose the generality, i.e., all possible local conserved quantities are shit invariant.
The approach shown in this Appendix is inspired by \cite{Chi24} and is simpler and more transparent than that presented in \cite{Shi19}.

The most general form of $k$-support local conserved quantity is expressed as
\eq{
Q=\sum_{l=1}^{k} \sum_{\bsA^l} \sum_{i=1}^L q_{\bsA^l,i}\bsA^l_i
}
where the coefficient $q_{\bsA^l,i}$ is position dependent.
We suppose that $k$ is the minimum number such that a nontrivial $k$-support local conserved quantity exists.
Our goal is to confirm $q_{\bsA^l,i}= q_{\bsA^l,i+1}$ for all $\bsA^l$ and all $i$.

This result is in fact readily shown in \sref{XYZh-step1} by examining the argument carefully.
We demonstrate it by taking a 4-support conserved quantity $Q$ and $\bsA^l=\x\y\z$ as an example.
As shown in \sref{XYZh-step1}, $\x\y\z$ and $\y\z\x$ form a pair.
Recalling the structure written in the column expression \eqref{doubling-pair}, we find that two coefficients $q_{\x\y\z, i}$ and $q_{\y\z\x, i+1}$ form a pair.
In this manner, we find pairs
\eq{
q_{\x\y\z, i}\lr q_{\y\z\x, i+1} \lr q_{\z\x\y, i+2} \lr q_{\x\y\x, i+3}\lr q_{\y\x\y, i+4}.
}
On the other hand, $q_{\y\x\y, i+4}$ also forms pairs as
\eq{
q_{\y\x\y, i+4} \lr q_{\z\y\x, i+3} \lr q_{\y\z\y, i+2} \lr q_{\x\y\z, i+1}.
}
It is easy to confirm that the sign and additional coefficients $J^{\bsA}$ of $q_{\x\y\z, i}$ and $q_{\x\y\z, i+1}$ are the same, which implies the desired result
\eq{
q_{\x\y\z, i}=q_{\x\y\z, i+1}.
}
In general, for $\overline{A^1\cdots A^{k-1}}$, by denoting two Pauli operators not equal to $A^{k-1}$ by $B$ and $C$, we have relations
\eq{
q_{\overline{A^1\cdots A^{k-1}},i}\lr q_{\overline{A^2\cdots A^{k-1}B},i+1}\lr q_{\overline{A^3\cdots A^{k-1}BC},i+2}\lr \cdots \lr q_{\overline{BCBC\cdots },i+k} \lr q_{\overline{CBC\cdots },i+k+1}
}
and
\eq{
q_{\overline{CBC\cdots },i+k+1} \lr q_{\overline{A^{k-1}CBC\cdots },i+k} \lr q_{\overline{A^{k-2}A^{k-1}CBC\cdots },i+k-1} \lr \cdots \lr q_{\overline{A^2\cdots A^{k-1}C},i+2}\lr q_{\overline{A^1\cdots A^{k-1}},i+1}.
}
These relations imply the desired result
\eq{
q_{\overline{A^1\cdots A^{k-1}},i}=q_{\overline{A^1\cdots A^{k-1}},i+1}.
}

In the case of the NNN-XYZ model, the above argument shows that a $k$-support operator $\bsA$ on site $i$ and site $i+2$ have the same coefficient.
To show that those on site $i$ and site $i+1$ are the same, we need to employ both cases where $k+2$-operators and $k+1$-operators are generated.
Let us consider $q_{\tx\ty\tz, i}$ as an example.
The above argument shows that $q_{\tx\ty\tz, i}=q_{\tx\ty\tz, i+2m}$ and $q_{\tx\ty\tz, i+1}=q_{\tx\ty\tz, i+2m+1}$ for any integer $m$.
By considering the case generating $k+1$-operators, we find a sequence of pairs
\eq{
q_{\tx\ty\tz, i}\lr q_{\ty\tz\x, i+2}\lr q_{\tz\x\ty, i+4} \lr q_{\x\ty\tz, i+6} \lr q_{\ty\tz\ty, i+7}.
}
In addition, we find a pair
\eq{
q_{\ty\tz\ty, i+7}\lr q_{\tx\ty\tz, i+5},
}
which implies the desired result
\eq{
q_{\tx\ty\tz, i}=q_{\tx\ty\tz, i+5}=q_{\tx\ty\tz, i+1}.
}
The generalization for general $\widetilde{A^1}\widetilde{A^2}\cdots \widetilde{A^{k/2-1}}$ is straightforward.

We finally exclude the possibility that a $k$-support conserved quantity $Q$ has shift-invariant $k$-support operators but non-shift-invariant $m$-support operators with $m<k$.
We prove it by contradiction.
Suppose that $k$ is the minimum number such that a nontrivial $k$-support conserved quantity $Q$ exists.
Consider $Q'=Q-TQT^{-1}$ with one-site shift operator $T$.
By construction, $Q'$ is conserved, and since $Q$ has shift-invariant $k$-support operators, $Q'$ is a less-than-$k$-support conserved quantity.
In addition, by expanding $Q'$ as $Q'=\sum_{l=1}^{k-1} \sum_{\bsA^l\in \calP^l} \sum_{i=1}^L q'_{\bsA^l,i}\bsA^l_i$, we easily see from the construction that $\sum_{i=1}^L q'_{\bsA^l,i}=0$ for any $\bsA^l$.
This directly implies that $Q'$ is not a trivial local conserved quantity (the Hamiltonian $H$ and the Z magnetic field in the case with symmetry), and hence $Q'$ is a nontrivial less-than-$k$-support conserved quantity, which is a contradiction.

\bigskip

{\it Acknowledgement}||
The author thanks Atsuo Kuniba, Yuuya Chiba, Mizuki Yamaguchi, and HaRu K Park for fruitful discussions.
The author also thanks Hosho Katsura for careful reading of the manuscript and various helpful comments.
The author thanks Hidehiro Saito for informing me various literature on zigzag spin chains.
The author was supported by JSPS KAKENHI Grants-in-Aid for Early-Career Scientists Grant Number JP19K14615.

\bigskip

{\it Conflict of interest}||
The author declares no competing interest.

\bigskip

{\it Data availability}||
 No datasets were generated or analyzed during the
current study.

\end{document}